\documentstyle[12pt,aaspp4]{article}

\received{1999 June 14}
\accepted{}
\journalid{}{}
\articleid{}{}

\slugcomment{Draft for ApJ}

\lefthead{Basri, Mohanty, Allard et al..}
\righthead{An Effective Temperature Scale for L Dwarfs from Resonance Absorption Lines of \cs and \rb}

\begin{document}
\def\cs{\ion{Cs}{1}\ }
\def\rb{\ion{Rb}{1}\ }
\def\na{\ion{Na}{1}\ }
\def\pot{\ion{K}{1}\ }
\def\hal{H$\alpha$\ }
\def \kms{km s$^{-1}$\ }
\def\msun{M$_\odot$\ }
\def\mj{M$_J$\ }
\def\teff{T$_{e\! f\! f}$~}
\def\vsini{{\it v}~sin{\it i}~}

\title{An Effective Temperature Scale for Late M and L Dwarfs, from Resonance Absorption Lines of \cs and \rb}

\author{Gibor Basri\altaffilmark{1}, Subhanjoy Mohanty\altaffilmark{1}, France Allard\altaffilmark{2}, Peter H. Hauschildt\altaffilmark{3},  Xavier Delfosse\altaffilmark{4,5}, Eduardo L. Mart\'\i n\altaffilmark{1}, Thierry Forveille\altaffilmark{4}, Bertrand Goldman\altaffilmark{6}}

\altaffiltext{1}{Astronomy Department, University of California,
    Berkeley, CA 94720.  basri@soleil.berkeley.edu; subu@astron.berkeley.edu; ege@popsicle.berkeley.edu}
\altaffiltext{2}{C.R.A.L (UML 5574) Ecole Normale Superieure, 69364 Lyon Cedex 7, France.  fallard@ens-lyon.fr}
\altaffiltext{3}{Dept. of Physics \& Astronomy \& Center for Simulational Physics, University of Georgia, \\Athens, GA 30602-2451.  yeti@hal.physast.uga.edu}
\altaffiltext{4}{Observatoire de Grenoble, 414 rue de la Piscine, Domaine Universitaire de St. Martin d'H\`eres, \\F-38041 Grenoble, France.  Thierry.Forveille@obs.ujf-grenoble.fr}
\altaffiltext{5}{Instituto de Astrof\'\i sica de Canarias, E-38200 La Laguna, Tenerife, Canary Islands, Spain.  delfosse@ll.iac.es}
\altaffiltext{6}{CEA, DSM, DAPNIA, Centre d'\'Etudes de Saclay, 91191 Gif-sur-Yvette Cedex, France.  bgoldman@cea.fr}

\begin{abstract} 

We present Keck HIRES spectra of 6 late-M dwarfs and 11 L dwarfs. Our goal is to assign effective temperatures to the objects using detailed atmospheric models and fine analysis of the alkali resonance absorption lines of \cs and \rb.  These yield mutually consistent results ($\pm$ 150 K) when we use ``cleared-dust'' models, which account for the removal of refractory species from the molecular states but do not include dust opacities.  We find a tendency for the \rb line to imply a slightly higher temperature, which we ascribe to an incomplete treatment of the overlying molecular opacities. The final \teff we adopt are based on the \cs fits alone, though the \rb fits support the \cs temperature sequence.  This work, in combination with results from the infrared, hints that dust in these atmospheres has settled out of the high atmosphere but is present in the deep photosphere. We also derive radial and rotational velocities for all the objects, finding that the previously discovered trend of rapid rotation for very low mass objects is quite pervasive.  To improve on our analysis, there is a clear need for better molecular line lists and a more detailed understanding of dust formation and dynamics.    
\end{abstract}



\section{Introduction}

M dwarfs are among the faintest and coolest stellar objects.  Their optical spectra are dominated by molecular band absorption.  The major opacity sources in the optical regions are TiO and VO bands, which produce a pseudo-continuum with the atomic lines superposed.  Recent surveys (e.g. DENIS, 2MASS) have turned up even cooler objects, including brown dwarfs (\cite{Delfosse99}; \cite{Kirkpatrick99a}).  The optical spectra of these objects are still characterized by molecular band absorption; however, the TiO and VO bands, which constitute the defining characteristic of the M spectral class, become very weak or are absent, while the metal hydrides begin to dominate.  This is presumably because of the depletion of heavy elements in their cool photospheres through the formation  (and perhaps subsequent gravitational settling) of dust grains, especially solid VO and perovskite (CaTiO$_3$) (\cite{Sharp90}; \cite{Tsuji96a}; \cite{Allard98}).  For this reason, a new spectral class, `L', has been proposed for these very cool low-mass objects (\cite{Martin97}; \cite{Kirkpatrick99a}; \cite{Martin99b}).  

To understand these very cool, low mass stellar and substellar objects, it is necessary to assign to them an effective temperature scale.  The construction of \teff sequences has been attempted in recent years by comparing the spectra of these objects to synthetic ones generated by atmospheric models.  Comparisons have been made on the basis of spectral energy distributions (SEDs) (\cite{Tinney98}; \cite{Ruiz97}) and infrared colors (\cite{Leggett98}). Attempts have also been made to construct a temperature scale using the molecular lines in the observed spectra, since these are sensitive to photospheric temperature. \cite{Tinney98} do this in the optical by ranking the objects in order of TiO, VO, CrH and FeH equivalent widths; \cite{Delfosse99} pursue a similar program in the infrared by using H$_2$O indices. \cite{Tokunaga99} do so with a spectral color index based on moderate dispersion spectroscopy in the K band.  \cite {Kirkpatrick99a} have performed a detailed analysis of low dispersion optical spectra as a first step towards defining the subclasses of the L spectral type. 

It is found that the models fit the observed SEDs and colors fairly well overall, and an effective temperature sequence is indeed derivable from the fits. However, as \teff decreases from early to late spectral types, some molecular bands which comprise most of the low resolution spectral features of these cool objects first increase and then decrease in strength due to dust formation (both through depletion from the gas phase and heating effects). The broadband infrared color indices (eg. J-K or I-K) saturate at very cool temperatures, and eventually also reverse direction (though it seems that I-Z or I-J might be monotonic with temperature). The molecular line lists in the models are not fully satisfactory, which could affect the temperature scale. It is therefore useful to have an alternative and complementary way of estimating effective temperatures.

We present such a method here: fitting synthetic spectra to the profiles of strong atomic alkali lines.  Their extremely low ionization potential allows the alkali metals (Li, Na, K, Cs and Rb) to remain neutral only at temperatures  below 3000K.  More importantly, their atomic resonance lines occur in the red and are thus prominent in very cool photospheres, displaying deep cores and broad wings in high-resolution spectra.  Finally and most importantly, the alkali metals are relatively undepleted by dust formation (\cite{Burrows99}), and their resonance lines are apparently formed in photospheric layers located above the region affected most by dust.  For these reasons, the resonance features of the alkali metals are excellent temperature indicators in cool photospheres, and grow monotonically in strength with decreasing temperature in the range of interest.  We therefore attempt to assign effective temperatures to 17 late-M and L dwarfs by fitting model spectra to their \cs and \rb high-resolution line-profiles.

In principle, the \na and \pot lines can be also be used for this purpose.  However,  we have not considered them in this paper, for the following reasons.  The \pot line in some of our sample objects is very broad, extending beyond one whole echelle order.  To fit models to it, neighboring orders must be patched together, a task fraught with sources of error and thus considered too risky for the fine-analysis we undertake here.  We also do not have observations of the \na lines for a number of our sample objects.  Therefore, we consider only the \cs and \rb lines in this paper.  We should point out that this method has its own weak points. The treatment of dust, or of the background molecular opacities around the lines, do influence the apparent strength of the line by changing the nearby continuum.  While it is unlikely that the molecular opacities are wildly off, we show that dust can have a major effect. 

The synthetic spectra we use are generated by the atmospheric models of Allard and Hauschildt (hereafter AH).  The models come with 3 treatments:  ``standard (no-dust)'', ``dusty'' and ``cleared-dust'' (discussed in \S3).  The first does not account for dust formation at all, while the second takes dust formation into account by considering both dust grain opacities and photospheric depletion of dust-forming elements (physically, this implies a haze of dust that is mixed throughout the atmosphere).  The third ignores dust opacities but accounts for photospheric depletion, simulating the state of affairs when dust forms and then gravitationally settles below the photosphere.  All 3 treatments treat line broadening in the same fashion.  We find that the ``standard'' and ``dusty'' models do a poor job of reproducing the optical observations -- the alkali lines in the first are too narrow and in the second they are too weak to fit most observations.  The ``cleared-dust'' models, however, fit the data much better. 

\section{Observations}
\subsection{Sample}

Until 1997, the only L-type object known was GD165B (\cite{Becklin88}), although it was not identified with that sobriquet. Indeed, a good spectral analysis of it has only recently appeared (\cite{Kirkpatrick99b}).  All the L dwarfs studied in this paper, therefore, are recent discoveries. They come primarily from the DENIS all-sky IR southern survey (\cite{Delfosse99}).  Some are from other sources.  Kelu-1 (\cite{Ruiz97}) was announced as the first confirmed field brown dwarf; it was found in a proper motion survey of faint red objects.  LP 944-20 is a late M star from the Luyten proper motion survey, which \cite{Tinney98} discovered contains lithium, making it a brown dwarf. G196-3B is a brown dwarf companion to a nearby young M star found by \cite{Rebolo98}; this is among the lowest mass brown dwarfs currently imaged.  LHS 102B is an L dwarf companion to another nearby M star, found by a proper motion study of the EROS project's images (\cite{Goldman99}). Recently, the 2MASS project has greatly added to the inventory of L stars (\cite{Kirkpatrick99a}).  Thus, our sample is not selected in any particular way, but is a first look at objects out of the initial discovery list of cool VLM field objects.  The list of objects observed is given in Table 1 (the spectral types listed in Table 1 are from the literature; see the end of \S5.1 for a discussion).  There are new objects being discovered monthly now by the DENIS, 2MASS, and Sloan surveys; it will soon be possible to study samples chosen in a more statistically meaningful way.

\subsection{Data Acquisition and Reduction}

Observations were made  with the W. M. Keck I 10-m telescope on Mauna Kea using the HIRES echelle spectrometer (\cite{Vogt94}).  The observation dates are listed in Table 1.  The instrument yielded 15 spectral orders from 640nm to 860nm (with gaps between orders), detected with a Tektronix $2048^2$ CCD.  The CCD pixels are 24$\mu m$ in size and were binned 2 $\times$ 2; the bins are hereafter referred to as individual ``pixels''.  Each pixel covered 0.1A, and use of slit decker ``D1'' gave a slit width of 1.15 arcsec projected on the sky, corresponding to 2 pixels on the CCD  and a spectral resolution of R = 31000.  The slit length is 14 arcsec, allowing excellent sky subtraction.  The CCD exhibited a dark count of $\sim$2 e-/h and a readout noise of 5 e-/pixel.   

The data were reduced in a standard fashion for echelle spectra, using IDL procedures.  This includes flat-fielding with a quartz lamp, order definition and extraction and sky subtraction.  The stellar slit function is found in the redmost order, and used
 to perform a weighted extraction in all orders.  The wavelength scale is established using a ThAr spectrum taken without moving the grating; the solution is a 2-D polynomial fit good to 0.3 pixels or better everywhere.  Cosmic-ray hits and other large noise-spikes were removed using an upward median correction routine, wherein points more than 7$\sigma$ above the median (calculated in 9-pixel bins) were discarded.  All points below the median were retained, to preserve the integrity of sharp absorption lines.  

For the purpose of comparing data to models, it is necessary to normalize both to the same continuum value.  However, the continuum in the data and the models is characterized by a large number of overlapping molecular lines; moreover, the data is not flux-calibrated, and has a superimposed echelle blaze function.  Thus, only a pseudo-continuum is derivable for the data and models.  A straight-line fit to the pseudo-continuum was obtained for each observed spectrum, and divided out for normalization.  The fit was made after manually discarding strong spectral lines (selected by eye), including the alkali absorption lines.  For \cs, the overall fit was obtained in the range 8500-8540A; for \rb, in the range 7930-7960A. 

\subsection{Radial Velocities}

Barycentric and stellar radial velocity corrections were derived for each observed spectrum. We thank G. Marcy for the IDL routine which calculates the barycentric correction. The radial velocities were derived by measuring the peak position for cross-correlation functions between each star and a radial velocity standard. We chose Gl 406 (M6) as our standard for the M dwarfs, and LHS 29224 (M9) and 2MASSW 1439+1929 (L1) as our standards for the L-dwarfs.

Gl 406 was chosen as a standard since its velocity has been given by several authors to good accuracy.  In particular, Delfosse has found it to be +19  \kms as part of his precision radial velocity project.  Measurement of the \cs line shift on an absolute wavelength scale for our December 1997 observation of Gl 406 yielded a velocity of +19 \kms.  A previous spectrum taken by Basri in November 1993 gives +18  \kms, and a new spectrum observed in June 1999 gives +19 \kms (using the December 1997 spectrum as calibrator).  However, measurement of the \cs line shift on an absolute wavelength scale yields a velocity of +43 \kms for a spectrum observed on June 2, 1997 (with an accuracy of about 1\kms).  Cross-correlation of this spectrum with the December 1997 Gl 406 observation also yields a velocity of +43 \kms.  Another spectrum of Gl 406 taken the next night (June 3, 1997), with a different grating setting, also yields a similar anomalous velocity (+40 \kms) upon cross-correlation with the December 1997 observation.  We examined the discrepant velocity of June 1997, comparing the ThAr spectra and night sky spectra with those of December 1997 to confirm that the grating had not moved from where we thought it was. There is a clear shift of the star compared with night sky lines on the same echelle frames for the two epochs. The stellar spectra themselves are otherwise almost identical. 

Radial velocities of all objects earlier than L0 were obtained by independent cross-correlation with the December 1997 and June 1999 Gl 406 spectra, in the spectral interval 8630 - 8753A~ (containing TiO, VO, CrH and FeH lines).  The results are mutually consistent.  We also obtained radial velocities of the objects earlier than L0 through cross-correlation with the June 1997 Gl406 spectra in the stated interval - the velocities obtained were consistent (within $\pm$5 \kms) with those obtained through cross-correlation with the December 1997 and June 1999 Gl406 spectra.  This implies that the anomalous Gl406 radial velocity of June 1997 is real.  Delfosse, however, has not seen radial velocity variations in his precision observations of Gl 406, including one only a few months away from those reported here. Either Gl 406 is a radial velocity variable, or we happened to observe a near twin quite close by (none is known). We have not discovered such variable behavior in any other of our sample objects.  However, \cite {TinneyReid98} obtained an ``inexplicable'' discrepancy in velocity for GJ 1111 compared to other published results. Such behavior could result if these are highly eccentric single-lined spectroscopic binaries. Clearly these systems bear closer monitoring.

Since the spectra of the objects L0 and later in our sample in the given interval are not very similar to that of Gl 406 (as TiO and VO disappear and CrH and FeH appear with decreasing temperature), cross-correlation for these objects was carried out using LHS 2924 (M9) and 2MASSW 1439+1929(L1) as independent standards.  Radial velocities were then calculated using the radial velocity of -37 \kms and -29 \kms obtained for LHS 2924 and 2MASSW 1439+1929 respectively by cross-correlation with the Gl 406 spectra.  The results were mutually consistent.  However, the very coolest objects (DENIS-P J0205-1159, 2MASSW 1632+1904 and DENIS-P J0255-4700; $\sim$ L5-L6), did not yield very good cross-correlation functions with either of the two calibrators.  Two factors probably contribute to this:  The lower S/N of the spectra of these three objects, and the fact that, at their low \teff, their spectral features begin to diverge from those of the two calibrators.

The radial velocity results are given in Table 1.  The stated errors in the Gl 406 December 1997 and June 1999 radial velocities are from comparing our values to those of Delfosse.  All other stated errors are calculated from the differences in radial velocity yielded by using different calibrator spectra.   

\section{Model Atmospheres}

As photospheric temperatures decrease, atoms combine to form molecules; upon still further cooling, atoms and molecules may coagulate to form dust grains.  Dust formation can change the atmospheric spectral characteristics in various ways.  For example, grains can warm the photosphere by backwarming, making the spectral distribution redder while weakening molecular lines (e.g., of H$_2$O and TiO) (\cite{Jones97}; \cite{Tsuji96b}).  Dust formation decreases the photospheric gas phase abundance of the atoms that form dust (e.g. Ti, Ca, Al, Mg, Si, Fe).  Furthermore, with  decreasing temperature, dust grains can become larger and gravitationally settle below the photosphere (\cite{Allard98}).  Thus, in any attempt to assign effective temperatures to cool photospheres, we must make use of models that take dust formation and behavior into account. Three basic models are considered:

{\it Standard (no-dust) Models}:  In these models, various molecules are formed with decreasing \teff, but no dust is allowed to form.  These are the NextGen series of models described by \cite{Allard96}. This scenario is known to be problematic  below a certain temperature; \cite{Tsuji96a} have argued that this temperature might be as high as 2800K.   A panel of spectra from these models in the \cs region is shown in Fig. \ref{stdmod}, left panel.  The opacity is dominated by TiO over most of the spectrum
 shown. There is also a band due to VO at 8528A~; this is treated with the JOLA (Just Overlapping Line Approximation) method, as no good line lists are available for this molecule. The apparent TiO strengths remain high throughout the temperature range (2500-1600K) shown, and the \cs line at 8521A~ grows moderately stronger.  It does not achieve anything like the observed strengths of \cs in L dwarfs.

{\it Dusty Models}:  Here, various molecules are formed with decreasing \teff, which then coalesce into dust grains that remain in the photosphere.  A range of grain sizes is assumed, and the resulting dust opacity taken into account, as well as the photospheric depletion of dust-forming atoms and molecules.   These models are described by \cite{Leggett98} and \cite{Allard98}.  The same spectral region as above is shown for this model approximation in Fig. \ref{stdmod}, right panel. Even at 2600K the spectrum looks rather different, as Ti is already condensing out and becoming unavailable to the molecular bands. A very different set of molecular lines is apparent at 2200K, and the \cs line has actually disappeared! This is not due to depletion of atomic cesium, but the growth of stronger opacity sources. At cooler temperatures, all the molecular lines are increasingly weakened by the growing dominance of the (featureless) dust opacity, leaving a rather smooth spectrum by 1600K. What is not apparent from these normalized plots is that the actual flux levels are quite different in the 2 panels. In the models without dust the TiO lines strongly suppress the flux, while the dusty case is brighter not only because of lower opacity but because the atmosphere is heated by absorption of radiation by the dust.

{\it Cleared-Dust Models}:  In this case as dust grains form they are assumed to gravitationally settle below the photosphere.  These models are constructed using the full dust equation of state (thus accounting for photospheric depletion of dust-forming materials) but neglecting dust opacity. \footnote {We have chosen the name ``cleared-dust'' models to denote the absence of dust {\it opacity} in these models.  In the simplest picture, formation of dust coupled with the absence of dust opacity could be accomplished by the gravitational settling of dust below the photosphere.  However, other scenarios may be invoked as well; for example, the dust may collect into photospheric clouds with a small covering fraction, or the dust may form very large grains with a small total blocking efficiency.  Thus the absence of dust opacity in the photosphere should not necessarily be taken to guarantee the absence of photospheric dust itself.}  

In these spectra (Fig. \ref{dustmod}, left panel) the strengths of both molecular and atomic lines grow at cooler temperatures.  Although TiO is gone, it is replaced by other molecular lines (such as CrH), but these are not nearly as strong in absolute opacity.  The VO JOLA features at 8528\AA~ grows stronger with cooler temperatures (not seen in the observations).  The \cs line becomes very strong at cooler temperatures in agreement with observations.  The line wings are easily able to overcome the rather weak molecular opacities, and the line is not only deep but broad.  These are collisional damping wings.  In Fig. \ref{dustmod} (right panel) we show the analogous spectra for the \rb region, with a VO JOLA feature at 7942\AA~. Similar behavior is apparent, except that the molecular opacities are even weaker, and the line comes to completely dominate the spectral region shown by 1600K.  This is partly due to the increased abundance of \rb compared to \cs (the \pot resonance lines, not shown, are vastly stronger than even \rb, due to a much higher abundance; these lines grow wings which cover more than 100 A).  It is the lack of TiO combined with the lack of dust opacities which allow the atomic lines to grow so strong in these models, and they resemble the observations most closely.

Naively, one might have expected the standard models to hold true in our hottest objects, the dusty models in the cooler objects, and the cleared-dust ones in the coolest objects.  However, our observations indicate that the dusty models are never representative of the optical observations.  On the other hand, \cite{Leggett98} and \cite{Goldman99} find their best fit to the infrared photometry (color {\it and} absolute flux) of L and very late M dwarfs with these dusty models.  One way of reconciling this disparity is to postulate that gravitational settling depletes dust in the upper photosphere (where the visible spectrum forms), but not in the lower photosphere (where the IR spectrum forms).  In reality, formation of dust clouds, dredging up of grains by convection, and various other meteorological phenomena must be modeled for an accurate portrayal of the situation.  In our present early stages of modeling, we expect the three models outlined above to at least bracket the observed spectra.

\section{Analysis}
In this section, we present a discussion of determination of rotational broadening for each star (\S4.1), model parameters used for profile-fitting (\S4.2), and the adjustments made to the models for purposes of comparison to the data (\S4.3). 

\subsection{Rotational Velocities}

Though the known rotation velocities of very early M-dwarfs ($<$M3) are usually $\lesssim$ 5 \kms, a significant fraction of the mid-M dwarfs (M3.5 - M6) show faster rotation (\cite{Delfosse98}).  Moreover, a trend toward increasing rotation velocity has been observed as one moves to later M spectral classes (\cite{Basri95}; \cite{Basri96}; \cite{TinneyReid98}).  Since stellar rotation (corrected for the inclination of the rotation axis) Doppler broadens the line profile, and changes the way it merges into the surrounding pseudo-continuum, rotational effects must be considered during profile-fitting.  We determine \vsini from molecular lines by the cross-correlation method described below.  Once the \vsini of the target objects are determined, the models are convolved with the corresponding rotation profiles.

For the spectral types earlier than L2 in our sample, \vsini is determined by correlating their spectra in the interval $\sim$8630 - 8753A~ with that of Gl 406 in the same interval.  This spectral interval contains molecular lines caused by VO, TiO, CrH and FeH.  The \vsini for Gl 406 has been determined by Delfosse et al. (1998) to be $<$ 3 \kms; thus it can be considered at the level of accuracy desired here to be non-rotating (since our instrumental broadening is of the same order).  

We first obtain the correlation function between our target spectrum and Gl 406 in the specified spectral interval.  We then artificially rotate our observed Gl 406 spectrum to various velocities by a convolution algorithm, and obtain the correlation function between each of the rotated Gl 406 spectra and the original unrotated Gl 406 spectrum, in the same spectral interval. Correlation functions are normalized to unity in the ``wings'', and additively adjusted so the amplitude of the main peaks are matched. The correlation function thus obtained that best matches that between the target object and the unrotated Gl 406 spectrum is selected, and the corresponding rotation velocity assigned to the target object.  

For the cooler L dwarfs (L2 and later) in our sample, Gl 406 can no longer be used as a calibrator.  Their spectral features in the chosen interval (or in any other interval available to us) are not very similar to those in Gl 406.  New molecular lines appear (as expected) with decreasing \teff, and the correlation function of Gl 406 with the L-dwarfs is no longer similar to those of Gl 406 with rotated versions of itself. It is therefore preferable to find a cooler calibrator.

An ideal calibrator should be a {\it non}-rotator, but we have not yet found any non-rotating L-dwarfs.  Both G196-3B and 2MASSW 1439+1929, the slowest L-dwarf rotators in our sample, have a \vsini of 10 \kms (as determined through cross-correlation with Gl 406).  We choose 2MASSW 1439+1929 as our calibrator, since it appears, from our model fits in Cesium and Rubidium, to be slightly cooler than G196-3B, and should thus be spectrally more similar to the later L-dwarfs, resulting in more robust correlation functions.

Through empirical tests, we determined that a rotating calibrator can indeed be used, so long as the targets rotate sufficiently faster than the calibrator. More precisely, we find that the cooler L-dwarfs need to be rotating faster than 30 \kms for their \vsini to be accurately determined (with errors of $\lesssim \pm$5 \kms), by cross-correlation with a 10 \kms calibrator. If the cooler L-dwarfs have vsini between 15 and 30 \kms, their rotational velocities can still be determined through cross-correlation with the 10 \kms calibrator, by applying a small, systematic correction factor of $\lesssim$ 5 \kms.  If the \vsini of the target is less than the \vsini of the calibrator, one can determine only that fact. The \vsini of all the objects in our sample of spectral type later than L1 have been determined by this method.  The values of \vsini for all our targets are listed in Table 2.  As noted previously in \S2.3, the coolest objects in our sample (DENIS-P J0205-1159, 2MASSW 1632+1904 and DENIS-P J0255-4700) do not yield very good correlation functions; the errors in their \vsini determination are thus somewhat higher.

\subsection{Model Parameters}

Effective temperature, surface gravity, metallicity and projected rotational velocity are all fundamental parameters which affect the observed spectra.  In our case  \teff is kept a free parameter, while \vsini is determined as described above.  Surface gravity and metallicity also affect the observed line-broadening.  Higher surface gravity increases the pressure at every atmospheric layer, causing increased collisional (Van der Waals) line-broadening, and a larger concentration of molecules at every layer.  The net effect is a broadening of the line with increasing gravity (\cite{Schweitzer96}).  

Metallicity enters in two opposite ways.  First, higher metallicity decreases the proportion of hydrogen particles (which are the main source of collisional broadening).  Second, higher metallicity also implies a decrease in pressure at a given optical depth (because of higher opacity), thereby reducing the line-broadening (this second effect probably dominates the first one, unless one goes to extremes of metallicity).  The net effect is a narrowing of the line with increasing metallicity (\cite{Schweitzer96}).  Higher metallicity also implies a greater abundance of alkali nuclei, which increases the strength of the atomic alkali lines. This effect is diminished to the extent that the alkali metals end up in molecules.  Notice that increasing gravity and increasing metallicity affect line-broadening in opposite senses; making the determination of gravity and metallicity a non-trivial task in cool dwarfs.  Metallicities for these relatively young objects may reasonably be supposed to be solar, but can vary somewhat about this value.  In this paper we assume a solar metallicity and defer dealing with metallicity variations.  

Stellar evolutionary models indicate that late-M and L dwarfs should have a surface gravity given approximately by log[$g$]=4.5 - 5.5 (\cite{Baraffe98}; \cite {Burrows97}).  Ideally one should keep $g$ a free parameter between these limits during profile-fitting.  At the time of writing, however, AH cleared-dust models were not yet available for values of log[$g$] other than 5.0. We therefore examined the no-dust AH models in the gravity range log[$g$]=4.5 - 5.5  and temperature range \teff = 2600 - 3200K (no-dust AH models with varying gravity at lower temperatures were also unavailable).  They show that the \rb line in this \teff range is saturated, while the \cs line is not.  However, both the \cs and \rb lines in most of our data (which is at much lower \teff than 2600K) are close to saturation.  Thus, we worked under the assumption that both \cs and \rb lines in our data (and in the cleared-dust models) behave similarly to the \rb line in the no-dust AH models.  With this assumption, we examined the behavior of the \rb line in the no-dust AH models, at log[$g$]=4.5,5.0 and 5.5.  We find that, at a given \teff, the log[$g$]=5.0 and log[$g$]=5.5 \rb lines are indistinguishable from each other, while the log[$g$]=4.5 line is much narrower.  Moreover, at a given gravity, the \rb line width is strongly dependent on \teff, becoming narrower with increasing \teff.  

What this means is that, with ``no-dust'' models, assuming log[$g$]=5.0 when it is really 5.5 will not lead to any significant errors in \teff determination.  On the other hand, if log[$g$] is really 4.5, then using 5.0 models will lead us to infer a {\it higher} \teff than is correct.  For example, in the ``no-dust'' models, we find that the model \rb line at \teff=2600K, log[$g$]=4.5 is very similar to the  model \rb line at \teff=2900K, log[$g$]=5.0.  Thus, if 4.5 were the correct value for log[$g$], using 5.0 would lead us to a \teff 300K higher than the real value.  Therefore, if indeed the \cs and \rb lines in the ``cleared-dust'' models at lower temperatures mimic the \rb line behavior in the higher temperature ``no-dust'' models, then our \teff are too high by $\sim$300K for low-gravity objects (see, for example, discussion of G196-3B in \S5.9).  On the other hand, at very low temperatures, we predict that the very high saturation in the resonance lines will lead to decreasing sensitivity to gravity in these lines.  Thus, in general, it appears that gravity variations lead us to \teff errors of $\lesssim$+300K in our sample.  This issue must be revisited once cleared-dust AH models at different gravities become available.  

There is an interplay between \teff and \vsini in cleared-dust AH models.  Decreasing \teff increases both line width and line depth, while increasing \vsini also increases line width but {\it decreases} line depth.  These effects are demonstrated in Fig. \ref{tvsini}.  We find that line width and line depth  considered together provide constraints on the effective temperature and rotational velocity.  We have used this fact to check the rotational velocities we derive through cross-correlation.  The results from profile fitting are consistent with those from cross-correlation.  However, since the cross-correlation method is more reliable and more precise, it is the method we use to determine \vsini.  We note also that, at the lowest temparatures in our data ($\sim$ 1700K), rotational broadening effects are largely overwhelmed by collisional broadening ones in the resonance lines (Fig. \ref{tvsini}).  Thus, the larger uncertainties in the \vsini of the coolest objects do not significantly affect their \teff determination.  
 
We must also ask whether collisional broadening, due to surface gravity or metallicity variations, could affect our correlation method results by providing broadening we mistake as rotation.  Models with different metallicities have not yet been produced, but no-dust AH models with different gravities are available.  By rotating models with different gravities (log[$g$] = 3.5, 4.5, 5.5) to various velocities and cross-correlating between them, we find that the cross-correlation function is insensitive to gravity, even over the large range of gravities considered.  Thus, the \vsini we derive are likely valid even if our sample objects do not have our assumed value of log[g]=5.0.  Insofar as metallicity effects are comparable to gravitational ones, the same should hold true for them.   

\subsection{Model Profile Adjustments}

As discussed, the models must first be corrected for rotational broadening effects (see \S4.2) through a convolution algorithm.  This effect occurs at the star.  There is also a finite instrumental resolution in the observations.  Thus, the rotationally broadened models were also broadened by Gaussians in accordance with the HIRES resolution of 31000 (this is an unimportant correction for most of the objects, given their high rotational velocities).  

In the data, the presence of overlapping molecular lines makes it necessary to renormalize each model with a derived pseudo-continuum (after an initial normalization with the appropriate blackbody spectrum).  Derivation of pseudo-continua in the model spectra is complicated by sudden strong breaks which are not apparent in the observed spectra.  These are produced by the Just Overlapping Line Approximation (JOLA) method used to treat VO, FeH, CaH, CaOH and NH$_3$ bands, for which insufficient parameters are currently available to allow a more accurate handling of the individual molecular lines. In the cleared-dust models the \cs line occurs outside (i.e., blueward of) the JOLA break region (Fig. \ref{dustmod}, left panel), while the \rb line occurs within
 a similar JOLA transition (redward of the break, Fig. \ref{dustmod}, right panel).  Thus, the models compute the \rb line with a background opacity higher than the real value, and we expect the model \rb fits to the data to be less accurate than the \cs fits.  This effect becomes smaller as the \rb line overwhelms the molecular opacities at the coolest temperatures.

We have compensated for the JOLA breaks in the following fashion.  For each resonance line, the model spectrum (in the range 8500-8540A~for \cs and 7930-7960A~for \rb) was divided into three sections - one blueward of the JOLA break  region (section 1), one redward of it (section 2) and a narrow section over which the break actually occurs (section 3).  A polynomial fit is obtained for section 1, after manually discarding strong spectral lines (selected by eye), and section  1 is normalized by this fit.  The flux in section 3 is assigned a single value -- the average, after normalization, of the 5 redmost points in section 1.  Section 2 is then separately normalized with a polynomial fit (again after discarding strong spectral features).  The combined effect of these procedures is to artificially remove the JOLA feature and concurrently normalize the model by the pseudo-continuum.  Note that these changes do not affect the actual calculation of the spectrum; they are applied post facto primarily for aesthetic reasons. They have the effect of making the entire model spectrum resemble the observations more closely. We emphasize that the profile-fitting we do is only in the line itself, while the normalization procedure detailed above affects primarily the spectrum outside the line. It should not affect our best fit parameters.

\section{Results}

\subsection{General Considerations for Profile Fitting}

Our preliminary profile-fitting of both the \cs and \rb lines indicates that the cleared-dust models are the only ones that fit the entire line data set well.  The dusty models do not agree with the data -- dust opacity in the models makes the lines much weaker than observed.  The no-dust models do not fit most of the data either, producing lines that are much narrower than those actually seen when the line depth is approximately correct, even after reasonable rotational broadening.  This is because in the no-dust models, the opacity in the wings of the resonance line is dominated by molecular opacity rather than the line wing opacity itself; hence only the narrow core of the resonance line is visible.  

When no-dust models do fit the data, the inferred temperatures (2400-2700K) agree with those from cleared-dust model fits.  Moreover, these inferred temperatures are high enough that we expect dust formation to be relatively unimportant; the no-dust and cleared-dust models should resemble each other.  In short, we found that cleared-dust models always produced fits at least as good as, and mostly much better than, either of the other two model types.  This is in agreement with the conclusions of \cite{Tinney98a}, who finds that cleared-dust models provide the best overall spectral shape fits in the optical regime for their sample of DENIS objects.  In what follows, therefore, we only consider the results of profile fitting with cleared-dust models.  We note again that in the infrared, dusty models do a better job of fitting SEDs (eg. \cite{Leggett98}). The left panels of the remaining Figures show our best fits for \cs and the right panels show the best fits for \rb.  When looking at the observed molecular regions, it may appear that the expected smoothing with increasing rotation is not present.  This is because the spectra are somewhat noisy, and noise spikes have the same ``sharpness'' regardless of rotation.  The eye is fooled by this, but a cross-correlation function is not.

Before going on to examine the results of profile-fitting for each of the objects in detail, certain general comments can be made.  First, the selection of best fits has been done by eye; any quantitative fitting procedure would, in our opinion, require more accurate synthetic spectra (especially for the molecular lines) and less noisy observations than are currently available.  Second, the figures show both data and models after smoothing with a 5-pixel box; this was done to allow the eye to follow the fit better by reducing noise fluctuations; a 5-pixel box, applied to the models and our current data sample, does this well without affecting the integrity of the resonance line significantly.  However, the actual model fits were derived by comparing  unsmoothed data and models, to eliminate the possibility of overlooking real but sharp features during profile fitting.

Third, there exist certain obvious general discrepancies between the models and the observed spectra, which are better noted at the outset rather than repeatedly on an object-by-object basis.  The first of these are poorly modeled or unmodeled molecular lines, that appear as additional broadenings in the observed \cs and \rb line wings, but not in the models (see for example the \cs and \rb fits to Kelu-1 (Fig. \ref{kelu1})).  In general, they appear to increase in number and strength with decreasing \teff.  For instance, there is no hint of them in the \cs wings of the Gl 406 data (\teff $\sim$2800K), while they appear strong and numerous in the observed \cs wings of DENIS-P J0255-4700 (\teff $\sim$1700K).  This is expected, as the resonance line wings become broader with decreasing \teff and include increasing numbers of the continuum molecular lines.  Hence, since the models do not exactly reproduce the observed continuum molecular lines, they also do not accurately reproduce the observed resonance lines at lower temperatures.  On the other hand, the rotational velocity of the objects also increases with decreasing \teff, which smooths out the molecular lines more than it does the stronger resonance lines.  At a given \teff, therefore, this allows better model fits to the data with increasing \vsini (compare the 25 \kms, 2200K \cs fit to DENIS-P J0909-0658 (Fig. \ref{d0909}), to the 10 \kms, 2200K \cs fit to DENIS-P J2146-2153 (Fig. \ref{d2146})).  Of course, some of the features within the resonance lines may simply be noise spikes and not real molecular lines, as is probably the case in our low S/N observations (e.g., \rb spectrum of DENIS-P J1228-1547 (Fig. \ref{den19})).  However, this is not likely to be the case for those features that repeatedly appear in the spectra of a number of different objects.  

Furthermore, for \rb, there is a break in the observed spectra at $\sim$ 7939A~that is not reproduced in the models; the lack of agreement between models and data is most pronounced for this feature in the \rb temperature range $\sim$ 2300-2700K (2MASSW 1439+1929 to Gl 406).  This feature, in fact, looks somewhat like the JOLA break that we labored to remove from the models.  However, a comparison between the uncorrected models (Fig. \ref{dustmod}, right panel) and the data in the 7930-7960\AA~ range shows that the observed feature occurs blueward of the JOLA break in the models, and is stronger than the JOLA breaks at the same temperature.  Also, the observed opacity break is limited to only a small span of wavelengths; this is not true for the modeled JOLA breaks, which persist over large wavelength spans.  Only in LP 944-20 (Fig. \ref{lp944}, right panel) does the observed opacity break resemble the (removed) JOLA break.  Finally, the model JOLA break persists to much lower temperatures than the observed feature.  Thus, even if this is the feature that the JOLA formulation tries to simulate, the simulation is not a good one.  Except for this particular feature, there is reasonable agreement between the models and the data in the 7930-7960\AA~ range after the model JOLA break has been corrected for, so we appear vindicated in making this correction.

The effect on the fine analysis of the background JOLA opacity in the model \rb is complicated. On the one hand, the extra background opacity will tend to make the line look weaker (similar to the effect of dust in the ``dusty'' models). This would require a cooler model to reproduce a given observed line strength, since the line grows with decreasing temperature. On the other hand, the effect of the JOLA break can be reduced by making the model hotter (which lets the model line appear stronger). The question is whether this reduction goes faster than the reduction in line strength caused by the hotter temperature. Our actual results suggest that it does, which leads us to derive generally hotter temperatures from the \rb line than the \cs line. This is why we base our final temperatures on \cs.

For \cs, there is an absorption line, centered on $\sim$8514A~, that appears in the models but is absent in the data.  This feature becomes weaker with decreasing effective temperature.  There is also a molecular feature at $\sim$ 8505\AA~, that becomes stronger with decreasing temperature in the models but follows the opposite trend in the data.  This leads to poor agreement between models and data at this wavelength at effective temperatures below $\sim$2000K. These features may both be due to water in the stellar atmosphere. 

There are other small anomalies in the \cs region. In all but one of the observed objects with \cs \teff between 2150 and 2800K (2MASSW 1439+1929 to Gl 406, except DENIS-P J1208+0149), there appear to be molecular bands (not strong in the models) at 8506 and 8516A; the latter impinges on the \cs line.  Thus, when the normalized models and data are overlaid, the models appear brighter than the data over this wavelength range.  To correct for this, the models (for only the objects indicated above) were also renormalized so as to fit the data between 8516 to 8524A. We show both normalizations in the figures for these objects, and tended to choose the temperature of the model renormalized to fit near the \cs line.  It is conceivable that this molecular band is the one that the model JOLA breaks try to simulate, but if so, the simulation is an inadequate one, for much the same reasons given above for the feature at 7939A.  It is worth noting that both features occur over the same temperature range, and might thus be caused by the same molecule, VO or TiO.

Finally, a note about which model fits are given, and why.  For spectra for which only a single model is given, no other model was found to reasonably fit the data.  Where two model fits are given, one model is a better fit to the resonance line core, while the other is a better fit to the resonance line wings.  

Using information on the effective temperatures found below which best fit our profiles,  \cite{Martin99b} have assigned subclasses to the L-type objects in our sample based on low dispersion spectral indices.  They propose a subclass designation similar but not identical to that proposed by \cite{Kirkpatrick99a}.  While hotter objects have the same classification in both schemes, \cite{Martin99b} extend to L6 for the coolest objects, while \cite{Kirkpatrick99a} extend to L8 for the same coolest objects.  Since \cite{Kirkpatrick99a} assign these objects a temperature of $\sim$1500K, while we find them to have \teff of $\sim$1700K, we prefer to follow \cite{Martin99b}. The designation "bd" in Tables 1 and 2 indicates that lithium has been seen in the object, certifying it as substellar (given the low temperatures). Its absence does not mean the object is definitely stellar; brown dwarfs above 60 \mj will eventually deplete lithium.

In Table 2 we provide a summary of our results.  The spectral types listed are from \cite{Martin99b} for the objects common to both our samples.  In most cases, their spectral classification is consistent with our \teff determinations (using the relation between spectral subclass and temperature given by them).  In the case of DENIS-P J1208+0149, however, \cite{Martin99b} find M9, while our \cs \teff belongs to a slightly later subclass (M9.5 in their scheme).  \cite {Martin99b} also assign L6 to DENIS-P J0255-4700 and 2MASSW 1632+1904, while our \teff determination suggests a slightly earlier type ($\sim$L5-5.5).  2MASSW 1146+2230, 2MASSW 1439+1929, DENIS-P J0021-4244 and DENIS-P J2146-2153 are not included in the \cite{Martin99b} sample.  In the cases of 2MASSW 1146+2230 and 2MASSW 1439+1929, therefore, the listed spectral types are from \cite {Kirkpatrick99a}, since they are consistent with our \teff determination.  DENIS-P J0021-4244 and DENIS-P J2146-2153 are not in the \cite{Kirkpatrick99a} sample either, so we assign a spectral type to them based on our derived \teff for the two objects and the spectral type given by \cite{Martin99b} for objects with similar \teff.

\subsection{Gl 406 (M6V)}

\cite{Delfosse98} find a \vsini of $<$3 \kms for Gl 406; at our level of accuracy, this is indistinguishable from no rotation at all.  With \vsini=0  \kms, we find \teff =2800K from the \cs cleared-dust model fits and 2700K from the \rb, in agreement with the result of \cite{Jones94}. The 2800K model \cs line is slightly shallower than what is observed (Fig. \ref{gl406}) but otherwise fits the line wings and core well.  The molecular lines near the resonance line are reasonably well fit after renormalization, while the fit to the molecular lines further away is good before renormalization. The model \rb line fits the observed line core and wings well, except in the partially modeled molecular feature in the right wing.  The molecular lines are also fit, and there is no significant worsening of the fit as one moves away from the line over the wavelength range considered here. It is noteworthy that this is the only object in our sample for which we derive a \cs \teff that is higher than the \rb one; in all our other objects, the \cs \teff is lower than the \rb. 
 
\subsection{LHS 2924 (M9V)}

We derive a rotational velocity of 10  \kms through cross-correlation with Gl 406.  An effective temperature of 2400-2500K is derived from the \cs fit, and 2600K from the \rb fit; these values are somewhat higher than the $\sim$2200K found by \cite{Jones94} from a low-resolution IR spectrum.  In both cases the line core and wings are fit very well (Fig. \ref{lhs2924}).  In Cesium, the molecular lines near the resonance line fit well after renormalization, while the lines further away match better before renormalization.  The partially modeled molecular lines are evident in the wings of the \rb line, but the overall fit to the molecular lines is good in Rubidium.

The effective temperature scale for late-M dwarfs is in fact still under discussion. These objects may be, in some ways, more sensitive to the treatment of dust than the L dwarfs (for which ignoring dust opacities altogether seems to work). We have a large sample of other late-M spectra, and will revisit this question in a subsequent paper (with updated AH models). The several objects listed as M9.5 below do not have exactly the same temperatures.

\subsection{LP 944-20 (M9V)}

This is a confirmed brown dwarf, with a mass between 0.056 and 0.064 \msun, aged between 475 and 650Myr (\cite{Tinney98a}).  Through cross-correlation with Gl 406, we derive a rotational velocity of 30 $\pm$ 2.5  \kms, which agrees with \cite{TinneyReid98}.  The \cs fit gives \teff =2400K and the \rb fit gives \teff =2600K. \cite{Kirkpatrick99a} assign a spectral type of M9V to LP 944-20, similar to LHS 2924. We find it marginally cooler than LHS 2924, at least as implied by the cesium fit.  In both \cs and \rb, the models fit the line core and wings well (Fig. \ref{lp944}), and the nearby molecular lines moderately well.  The higher rotational velocity smooths the model molecular lines to a degree comparable to that observed, but there is poor agreement between individual lines in the models and the data.  

\subsection{DENIS-P J0021-4244 (M9.5V)}

We obtain a \vsini of 17.5$\pm$2.5 \kms for this object, through cross-correlation with Gl 406.  The \cs fit yields an effective temperature of 2300K, and the \rb fit an effective temperature of 2400-2500K.  The resonance lines in both cases are fairly well fit (Fig. \ref{d0021}), with the data being slightly wider than the models in the wings, due to poorly modeled molecular lines.  In \rb, the 2400K model is a little deeper than the data and overestimates the width of the line by fitting the unmodeled molecular line on the right, while the 2500K line matches the true width of the line but is slightly shallower than the data.  The nearby molecular lines are well modeled in the \rb section (at least in shape, if not in depth) and somewhat less so in the \cs.  The lines further away from the resonance line are well reproduced in the \cs section before renormalization.  The TiO features in this object are quite similar to those of BRI 0021 (M9.5).

\subsection{DENIS-P J1208+0149 (M9V)}

We derive \vsini = 10 \kms for DENIS-P J1208+0149, through cross-correlation with Gl 406.  An effective temperature of 2200-2300K is found from the \cs fits, and of 2500K from the \rb fits.  The \rb fit is good, except for the unmodeled molecular lines in the wings.  The fit to the molecular continuum in this spectrum is good as well.  In \cs, the line core is fit well by the 2200 and 2300K models, though the former is slightly broader and the latter a little shallower than the data, but both models appear much stronger than the data at the right edge of the resonance line.

Our results are slightly puzzling. \cite{Martin99b} find this object to be of spectral class M9, while our \cs \teff puts it at $\sim$ M9.5. However, our \rb \teff, which agrees with the \cs \teff ordering for most of the objects in our sample, does not in this case but implies instead that the object is somewhat hotter than 2200-2300K. Unfortunately, we do not have color data for this object, which might help resolve these discrepancies. The opacity band over 8516-8524A~ which is apparent in all other objects in the temperature range 2150-2800 K is not seen in this object.  These anomalies may result from the noisiness of the data.  This will have to be resolved with better spectra and improved models. 

\subsection{DENIS-P J2146-2153 (L0)}

We find a rotational velocity of 10 $\pm$ 2.5  \kms for DENIS-PJ2146-2153, through cross-correlation with Gl 406.  An effective temperature of 2200K is derived from the \cs fit and of 2400K from the \rb fit.  In the \cs section, the inclusion of unmodeled molecular lines makes the observed resonance line appear broader than the model in the wings.  The \rb resonance line is well modeled, though the data is slightly broader than the model in the left wing, due to a partially modeled molecular line (Fig. \ref{d2146}).  Both models are good fits to the continuum molecular lines, the Cesium models slightly better so before renormalization than after.  

We also note that while we assign this object to class M9.5, the low resolution spectrum of \cite{Tinney98} indicates M9; in particular, at low-resolution this object appears hotter than BRI 0021 (M9.5). A comparison of the TiO features in our spectra with a HIRES spectrum of that object bears this out. We need the \cs spectrum of BRI 0021 to sort this out; our \rb line here is stronger than that reported for BRI 0021 by \cite{Basri95}.

\subsection{DENIS-P J0909-0658 (L0)}

We obtain a rotational velocity of 25$\pm$2.5 \kms for this object, through cross-correlation with Gl 406.  An effective temperature of 2200K is derived from the \cs fit, and 2400K from the \rb fit.  In both cases, the line core and wings are well simulated by the models (Fig. \ref{d0909}), except a poorly modeled molecular line in the right \rb wing.  The \cs line does not show any evidence of unmodeled molecular lines; this may be a result of this object's higher \vsini, which would tend to smooth out such lines, or the fact that the previously unmodeled molecular line has disappeared with decreasing \teff.  The former explanation would seem to be the more correct one, since 2MASSW 1439+1929, which we find (see below) has a very similar \cs \teff ($\sim$2150K) but is a slower rotator (10 \kms) does show evidence for unmodeled lines in \cs.    

Nearby molecular lines are fit well in the \cs section, and poorly in the \rb section, which appears noisier than the \cs section.  The low resolution spectra of this object also suggest it is a good marker of the start of the L spectral type (\cite{Martin99b}).  

\subsection{G196-3B (L1)}

This sub-stellar object was recently discovered by direct imaging (\cite{Rebolo98}), at a distance of $\sim$ 300 AU around the young low-mass star G196-3 (M3Ve).  Its mass has been determined to be 25 (+15/-10) \mj and its age to be $\sim$ 100Myr (from the activity level of the primary star).  By comparing low-resolution spectra of G196-3B with those of Kelu-1 and DENIS objects, Rebolo et al. infer an effective temperature of 1800-2000K.

We derive a rotational velocity of 10$\pm$2.5  \kms for G196-3B through cross-correlation with Gl 406. This is relatively slow at such low mass, especially for such a young object (however, this may be simply a projection effect). This gives us an effective temperature of 2200K from the \cs model fit and 2400K from the \rb fit.  The fit in \cs is good in the line core and wings, except at the very edge of the red wing.  The overall fit to the molecular lines in the Cesium section is good before renormalization.  The model \rb line is slightly deeper than the data; the observed core may be chopped by noise.  It is also slightly narrower than the observation in the wings due to poorly modeled molecular lines.  The overall fit to the molecular lines is good in the Rubidium section, though the data appears somewhat noisy.

Our derived \teff is higher than Rebolo et al.'s estimate.  However, the molecular band between 8516A~ and 8524A~ that is present in all other objects in our sample that have \teff $\gtrsim$ 2200K, is present in G196-3B as well, giving credence to our higher temperature estimate.  On the other hand, the low-resolution spectrum obtained by Rebolo et al shows a clear lack of TiO bands and only a very faint VO band, indicating that this is an early L class object.  

Our high effective temperature might be due to effects of low surface gravity.  Given the apparent youth of G196-3B, it is reasonable that the object is still contracting. As mentioned in \S4.2, an analysis of ``no-dust'' AH models indicates that attempting to fit low-gravity scenarios with a higher gravity model might lead to spuriously high \teff values.  This question will have to be examined in the future with newer models.  Since the no-dust AH models also indicate that the TiO band ($\sim$7040-7140A) is relatively insensitive to gravity but sensitive to temperature, it may also be useful in separating gravity and temperature effects in G196-3B, which shows a lack of TiO bands. 

\subsection{2MASSW 1439+1929 (L1)}

This is a very high proper motion object, at a distance of $\sim$15.1 pc (\cite{Kirkpatrick99a}).  We find a rotational velocity of 10$\pm$2.5 \kms for this object through cross-correlation with Gl 406.  The cross-correlation function was checked against that obtained for G196-3B - since both objects are found to have \vsini of 10 \kms through cross-correlation with Gl 406, and are of similar spectral type, their correlation functions with Gl 406 should be identical - and \vsini $\sim$ 10 \kms was confirmed.  This object was subsequently used as a calibrator for deriving the radial and rotational velocities of all the later L dwarfs in our sample.  

The \cs model fits imply an effective temperature of 2100-2200K for this 2MASSW 1439+1929, and the \rb fits 2300K.  Both 2100 and 2200K models are slightly narrower than the data in the red wing of \cs, because of unmodeled molecular lines.  The 2100K model also seems to reproduce the molecular line in the blue wing, but is slightly deeper than the data, while the 2300K model does does not reproduce the observed blue wing as well, but matches the core depth better.  The molecular lines near the \cs line are reasonably matched by both models after renormalization, and the general molecular continuum well fit before renormalization.  This is the coolest object in which the anomalous opacity band between 8516-8524A~ is seen.  In the Rubidium section, the model is slightly narrower than the observed \rb line due to unmodeled molecular lines, and the continuum is generally well matched in shape but not strength. Our spectral classification of 2MASSW 1439+1929 as L1 agrees with result of \cite{Kirkpatrick99a}.    

\subsection{Kelu-1 (L2)}

Kelu-1 is a free-floating brown dwarf (\cite{Ruiz97}) at a distance of $\sim$ 20pc (Ruiz: priv. comm.).  A rotational velocity of 60$\pm$5  \kms is derived through cross-correlation with 2MASSW 1439+1929. From the \cs fits an effective temperature of 2000K, and from the \rb fits, a temperature of 2200K are found.  In \cs, the inclusion of poorly modeled lines makes the data appear much broader than the model.  In \rb, again, the data appears broader than the best-fit 2200K model.  The cores of the observed \cs and \rb lines appear to be somewhat chopped by noise.  We note here that the \cs and \rb fits, at 2000 and 2200K respectively, are substantially better if a \vsini of 80 \kms is used, instead of 60 \kms.  However, the higher velocity is not supported by our cross-correlation \vsini determination.

Our derived rotational velocity for Kelu-1 is very high, even given the trend towards increasing rotational velocity as one moves to the bottom of the main sequence (\cite{Basri96}). Presumably we are near its equatorial plane.  In both \cs and \rb, the models match the overall observed smoothness of the molecular lines, lending credence to our high derived \vsini, but do not match the lines in detail.  If indeed our derived \teff (which is dependent on our derived \vsini) for Kelu-1 is correct, then we find it to be cooler than 2MASSW 1439+1929 but hotter than 2MASSW 1146+2230, which is in agreement with the results of \cite{Kirkpatrick99a}.  Moreover, \cite{Ruiz97} find that the best fit to the Kelu-1 spectral energy distribution is given by a dusty AH model with \teff = 1900 $\pm$ 100 K and log[g] = 5.0 - 5.5 (with [M/H] fixed at 0.0).  \cite{Leggett98} find a best fit to the Kelu-1 IR colors (using AH dusty models) at \teff $\lesssim$ 2000K, and [M/H] = 0.0 (holding log[g] fixed at 5.0).  Together, the two studies suggest that log[g] = 5.0-5.5 and [M/H] $\sim$ 0.0, consistent with the values we have chosen, are suitable choices for Kelu-1, and that our \teff is also in the right ballpark.  Taken together, this again suggests that our \vsini is also approximately correct.      

One implication of such a high \vsini is that spectra with an exposure time of an hour or more actually sample most of the stellar surface.  Searches for variability or ``weather'' on this object, must therefore be done using short exposures (forcing low spectral resolution).  The rapid rotation may well be the explanation for the rather different equivalent widths reported by different authors (\cite{Rebolo98}).  The rotation means that the pseudo-continuum will be set somewhat differently at low, medium, and high spectral resolutions.  In fact, we now have high resolution spectra of Kelu-1 from 4 separate nights spanning a year (2 pairs), and see no evidence of variability in the line profiles.  

We also note here that a parallax for Kelu-1 has been found (Ruiz; private communication); this parallax indicates that Kelu-1 appears somewhat brighter than would be expected from \teff $\sim$ 1900K (as advocated by \cite{Ruiz97}).  This may imply that this object is a binary (although it appears single in HST images).  
   
\subsection{DENIS-P J1058-1548 (L2.5)}

A corrected rotational velocity of 37.5$\pm$2.5 \kms is obtained by correlation with 2MASSW 1439+1929.  An effective temperature of 1900-2000K is derived from the \cs fit, and of 2000K from the \rb fit, in agreement with the value ($\sim$ 2000K) derived by \cite{Leggett98} from infrared photometry for [M/H]=0.0. In \cs, the 1900K model matches the wings of the line but is slightly deeper than the data, and the 2000K model matches the depth of the line but is slightly narrower than the data (Fig. \ref{d1058}).   In \rb, the model is both slightly deeper and narrower than the data; this may be due to a core that is chopped by noise and poorly modeled lines in the wings.  The pseudo-continuum is not as smooth as the models, perhaps because of noisy data.

\subsection{2MASSW 1146+2230 (L3)}

This object is a close double (separation $\sim$1''); an LRIS spectrum reveals the companion to be a background star of much earlier type (\cite{Kirkpatrick99a}).  2MASSW 1146+2230 shows a strong lithium line with an equivalent width of 5.1A~ (\cite{Kirkpatrick99a}). We derive a \vsini of 32.5$\pm$2.5 \kms for this object through cross-correlation with 2MASSW 1439+1929.  The \cs model fits give a \teff of 1900-2000K, while the \rb fits give 2100K.  The observed spectrum appears noisy, and the models match the general trend but not the observed sharpness of the continuum lines in both the Cesium and Rubidium sections.  The \cs line is chopped in the very core by noise.  

This is one of our only objects in which the ordering implied by the \teff from \rb fits does not agree with that implied by the \teff from \cs.  The \rb fits imply that 2MASSW 1146+2230 is slightly hotter than DENIS-P J1058-1548, while th \cs fits imply that it is slightly cooler.  However, the two objects are very close in spectral type and \vsini, so, given the noise in our spectra, this does not constitute a serious anomaly.  If we follow our convention of giving the \teff derived from \cs more weight than that from \rb, we find 2MASSW 1146+2230 to be of spectral type L3, in agreement with the result of \cite{Kirkpatrick99a}.    

\subsection{LHS 102B (L4)}

LHS 102B is an L dwarf proper motion companion to LHS 102 (GJ 1001), a field mid-M dwarf.  It was very recently discovered in the course of the EROS 2 proper motion survey (\cite{Goldman99}).  The rotational velocity of this object is found to be 32.5$\pm$2.5 \kms, from cross-correlation with 2MASSW 1439+1929.  From \cs model fits, we find a \teff of 1800-1900K and from \rb 1900K. In both \cs and \rb, the resonance lines are well modeled (Fig. \ref{lhs102b}), though the models are slightly narrower than the data, because of poorly modeled molecular lines.  The \cs fit is also somewhat deeper than the data.  The overall fit to the molecular lines is better in Cesium than in Rubidium; in both cases (though more so in Rubidium) the model continuum is smoother than the data, probably due to noisy data.  

This object is similar to GD165B in general appearance.  It shares with GD165B an ambiguity about whether it is really a brown dwarf. While the lithium test has not been applied to GD165B (\cite{Kirkpatrick99b} have argued persuasively that it will fail); LHS102B definitely fails the lithium test. This does not mean that either object is stellar -- only that they are  older than about 200Myr and greater in mass than 60 \mj. It is likely that objects at about this temperature are near the minimum main sequence temperature. When the L subclass of the minimum main sequence temperature (for a given metallicity) is precisely identified, all cooler objects can safely be certified as brown dwarfs without regard to the lithium test.

\subsection{DENIS-P J1228-1547 (L4.5)}

This was the second field brown dwarf to be confirmed by the lithium test (\cite{Martin97}). The corrected rotational velocity for this object was found to be $\sim$22$\pm$2.5 \kms through cross-correlation with 2MASSW 1439+1929.  From \cs model fits, a \teff of 1800K is found, and from \rb fits, 1900K.  However, the \rb data is very noisy, and thus the effective temperature derived from it is rather approximate.  The \cs line is well modeled (Fig. \ref{den19}), but the model is slightly narrower than the data in the wings, due to poorly modeled molecular lines, and the data is chopped in the very core by noise.  The continuum in the Cesium data also appears rather noisy, and does not match the model smoothness.  However, recently \cite{Martin99a} have found that this object is actually a binary (separation 0.3 arcsec), with nearly equal brightness. It may be that one brown dwarf is slightly cooler than the other, in which case we are looking at a composite spectrum, and the issues of line-width and noisiness just mentioned will have to be reexamined after taking binarity into account.  \cite{Leggett98} derive an approximate effective temperature of 2000K (with [M/H]=0.0) for DENIS-P J1228-1547 from infrared photometry, and find it to be cooler than DENIS-P J1058-1548, in agreement with our results (and that of \cite{Tokunaga99}). 

\subsection{DENIS-P J0205-1159 (L5)}

A corrected rotational velocity of $\sim$22$\pm$5  \kms was derived for DENIS-P J0205-1159 through cross-correlation method with 2MASSW 1439+1929.  From the \cs fits we get a \teff of 1700-1800K.  The noisiness of the \rb data precluded diagnostic profile-fitting to the \rb line; however, the 1700K model generally matches the form of the data.  The \cs line is well modeled, though somewhat noisy as well - the very core of the line appears chopped by noise , and the molecular continuum in the data does not match the model smoothness (Fig. \ref{d0205}). Unmodeled features in the wings of the observed \cs line may be due to noise or poorly modeled molecular lines.  Both models exclude the apparently real molecular line in the right wing, but the 1700K model fit assumes that the features in the left wing are noise, while the 1800K fit assumes that they are real and thus excludes them completely.  

\cite{Leggett98} use infrared photometry to propose an effective temperature of  $\sim$ 2000K, which is higher than our result. More discrepantly, they find it to be hotter than both DENIS-P J1228-1547 and DENIS-P J1058-1548, which is at definitely at odds with our results. \cite{Tinney98} and \cite{Delfosse99}, on the other hand, find the same ordering in temperature based on low dispersion spectra as we do. \cite{Tokunaga99} find the same from moderate resolution K band spectra (and a new proposed K color index diagnostic ratio). These discrepancies show why it is valuable to have several independent temperature calibration methods. We suspect this object is cool enough to be classified as a brown dwarf on the basis of temperature alone.

\subsection{2MASSW 1632+1904 (L6)}
2MASSW 1632+1904 is the coolest L dwarf in \cite{Kirkpatrick99a}, and they find a possible lithium line at their detection limit (EW $\leq$ 9.4A~). We do not have the lithium line in our observations of this object. We derive a \vsini of 30$\pm$10 \kms for this object through cross-correlation with 2MASSW 1439+1929.    

Our spectrum for this object is noisy.  From the \cs model fits we derive an effective temperature of $\sim$1700K.  The model matches the general trend but not the sharpness of the lines in the noisy continuum.  In the Rubidium section, the noisiness of the data precluded any accurate fits, but both 1600 and 1700K models appear to follow the general shape of the \rb line.  

\cite{Martin99b} ascribe a spectral type of L6 to this object; we have adopted this since it is consistent with our \teff determination.  \cite{Kirkpatrick99a}, on the other hand, classify this object as L8.  It is to be noted, however, that they correspondingly find 2MASSW 1632+1904 to have a \teff of $\sim$1500K.  A \teff of $\sim$1600-1700K, as we find, corresponds to $\sim$L6 in their classification scheme as well.  It is also possible that our derived \teff is higher than that of \cite{Kirkpatrick99a} because this is a lower-gravity object (as discussed in \S4.2.  ``Cleared-dust'' models at varying gravities are needed to settle this question.   

\subsection{DENIS-P J0255-4700 (L6)}
This is our object with the strongest \cs line, so we think it is the coolest of the sample. For this object the rotational velocity was found to be 40$\pm$10  \kms by cross-correlation with 2MASSW 1439+1929.  From \cs model fits, we find a \teff of 1700K, and from \rb, 1900K. In \cs, the inclusion of unmodeled molecular lines makes the smoothed data appear much broader than the model.  The observed molecular continuum appears somewhat noisy, but seems to match the overall smoothness of the models.  The \rb data is noisy, but the resonance line is reasonably fit by both 1900K after excluding the unmodeled molecular lines. 

Despite the fact that it does not show lithium, it is very likely a brown dwarf based on its \cs temperature (with mass greater than 60 $M_{J}$).  \cite{Griffith98} worry about whether lithium may not show up due to formation of molecules such as LiCl, but this object is not too much cooler than DENIS-P J1228-1547 (which shows lithium). Even more reassuring is the detection of lithium in some objects this cool by \cite{Kirkpatrick99a}, although they do see a possible beginning of the expected weakening of lithium at very cool temperatures. The results of \cite{Griffith98} are suspect, since they have not done a self-consistent treatment of the stellar model atmosphere. Our models include lithium molecular formation but predict that strong lithium lines remain at these temperatures.

\section{Conclusions}

In general we find that the cleared-dust model profiles fit the observed \cs and \rb resonance lines reasonably well (with errors of $\pm$ 50K in the fitting).  We further find that, in a given object, the effective temperatures derived independently from \cs and \rb agree to within $\pm$ 150K, with the \rb fit consistently giving the higher temperature.  We think this is due to the JOLA modeling of overlying molecular opacities, and the overlap of a JOLA band with the \rb line.  We find that the \cs and \rb analyses independently imply the same ordering of objects by effective temperature. The temperature scale derived from these lines may be a little hotter than that which seems to be coming out of consideration of low dispersion spectral energy distributions. This could be due to our approximate treatment of dust or variations in gravity and/or metallicity, or perhaps the infrared treatment needs adjustment. Our final temperatures are based on the \cs lines.  

On the whole, we find good agreement between the models and the data with log[g]=5.0 and [M/H]=0.0.  With these values, we estimate our errors in \teff to be $\pm$50K, since our model grid has a spacing of 100K.  However, an analysis of ``no-dust'' AH models indicates that gravity variations in the log[$g$]=4.5-5.5 regime may lead us to infer a \teff greater by $\lesssim$ 300K than the real value, for low-gravity objects.  Metallicity effects, which act in an opposite sense to gravity ones, may also affect our \teff values.  ``Cleared-dust'' models with varying gravities and metallicites are needed to resolve this issue.  

In the cases where the models do not reproduce the resonance lines very well, various effects may be responsible - collisional broadening effects due to metallicity and/or surface gravity variations, imperfectly modeled molecular opacity overlapping the line, or a low S/N ratio in the data.  We find that the molecular lines are not modeled as well as the resonance lines are, and that the fit to the molecular lines generally worsens as one moves lower in effective temperature.   Both effects are expected, given the comparative paucity of well-determined parameters for many of the observed molecules.  Obviously it is desirable to improve our modeling of these lines. These affect the analysis of the objects at both high and low spectral dispersions (and in both the optical and near-IR).  

The treatment of dust is still problematic.  Although it must be true that there is rather little dust opacity in the far red (or we would not observe such strong atomic lines), it is not clear that our assumption of $no$ opacity is valid. We must also reconcile the results in the far red with those in the near infrared, which apparently require more dust opacity. We suggest that the dust has largely settled out of the portion of the atmosphere sampled by the alkali resonance lines, but is present in the lower photosphere where the near infrared is formed. \cite{Tsuji99} have made a similar suggestion in the context of Gl 229B.  It will be important to study line profiles in the near infrared to help sort this out. The suggestion is that stratified dust models are worth pursuing. At the moment, the height or extent of such dust stratification can probably be better informed by observations than theory.  This is the weak point in our analysis; the addition of some dust opacity would tend to reduce the inferred temperatures. It is not clear whether the good fits to the line shapes can be preserved as dust opacity is added (and a significant amount is needed to affect these very strong lines). 

The temperature scale found here led \cite{Martin99b} to propose a subclass designation scheme for the L spectral class in which L0 occurs at about 2200K, and each subclass is 100K cooler. For the coolest L dwarf in common with \cite{Kirkpatrick99a} (2MASSW 1632+1904), there is a disagreement in temperature. We predict that there is still a gap of a few hundred degrees between such objects and the T dwarfs, while \cite{Kirkpatrick99a} believe the gap is quite small. On the other hand, variations in gravity and/or metallicity may be causing us to infer an artificially high \teff for this object.  Depending on the resolution of this issue, either the \cite{Martin99b} or \cite{Kirkpatrick99a} L star classification scheme should be modified in order that the L stars extend down to the appearance of methane in the K band.

One clear and fairly model independent result from our analysis is that the average rotation velocity of very low mass stars gets higher and higher as one moves down through the bottom of the main sequence. It is clear that as hydrogen burning begins to turn off near the substellar boundary, the magnetic braking which affects all higher mass convective stars is also weakening. Our sample is entirely from the field, and some objects are several hundred Myr old (although there is certainly an observational bias against finding very old objects, particularly if they are brown dwarfs). It appears that there is relatively little angular momentum evolution among these objects, and that they are typically born with relatively rapid rotation. Their lack of magnetic activity is seen directly (through a lack of \hal emission), as well as indirectly in their rapid rotation. The effort to understand these very low mass objects has just begun.

\acknowledgments

{\it Acknowledgments}: 
This research is based on data collected at the  W.~M. Keck Observatory, which is operated jointly by the University of  California and California Institute of Technology. GB and SM acknowledge the support of NSF through grant AST96-18439. SM would like to thank GB and Don McCarthy (Steward Observatory) for invaluable mentorship.  EM acknowledges support from the Fullbright-DGES program of the Spanish ministry of Education. AH's work is supported by the CNRS, a NASA LTSA NAG5-3435 and a NASA EPSCoR grant to Wichita State University.  Some of the calculations presented in this paper were performed on the IBM SP2 of the CNUSC in Montpellier, and at the Cray T3E of the CEA in Grenoble.  PHH's work was supported in part by NSF grant AST-9720704, NASA ATP grant NAG 5-3018 and LTSA grant NAG 5-3619 to the University of Georgia.  Some of the calculations presented in this paper were performed on the IBM SP2 and the SGI Origin 2000 of the UGA UCNS and on the IBM SP of the San Diego Supercomputer Center (SDSC), with support from the National Science Foundation, and on the Cray T3E of the NERSC with support from the DoE.  We thank all these institutions for a generous allocation of computer time.

\newpage

\plotone{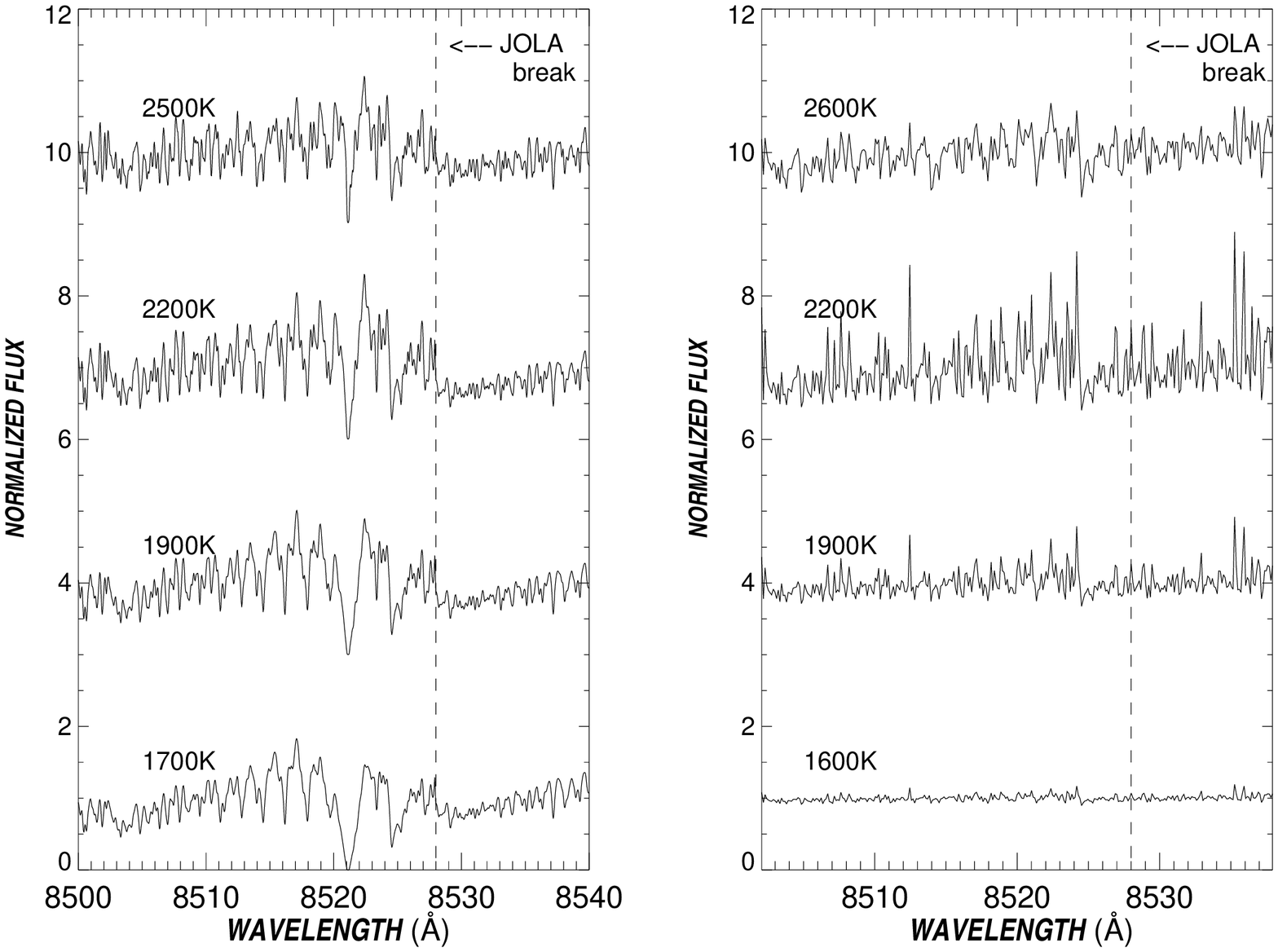}
\figcaption{\label{stdmod} {\it Left panel}: Standard (no-dust) models for \cs, with the VO JOLA break at 8528\AA~ indicated.  Molecular opacity allows only the core of the resonance line to be seen.  {\it Right panel}: Dusty models for \cs.  Veiling due to dust makes the resonance line appear very weak. The molecular features have changed due to depletion of elements into the dust phase. Each model is offset upwards by three from the lower model.  In these figures and all  following ones, log[g]=5.0 and [M/H]=0.0 }

\plotone{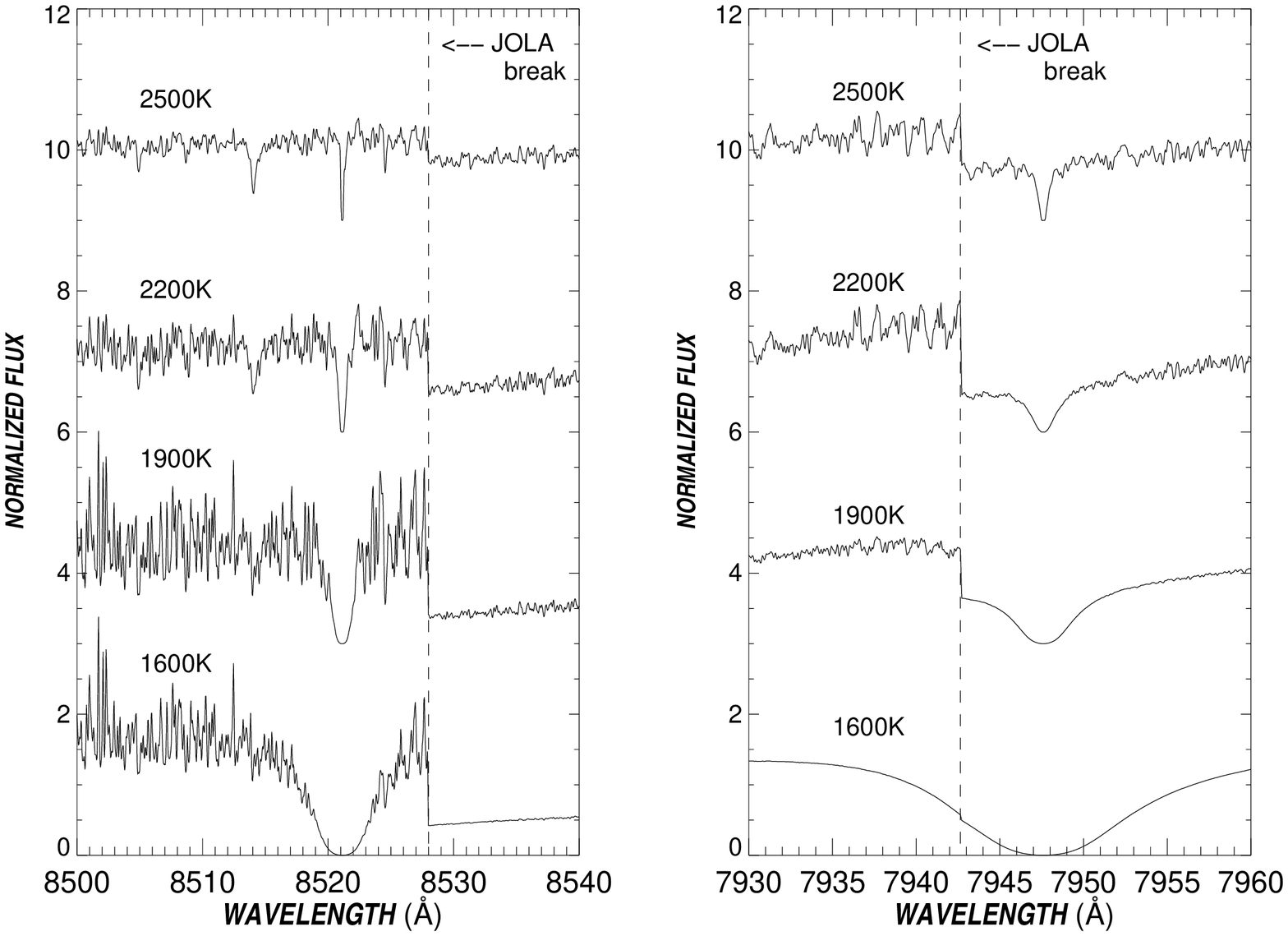}
\figcaption{\label{dustmod} {\it Left panel}: Cleared-dust models for \cs.    {\it Right panel}:  cleared-dust models for \rb, with the VO JOLA break at 7942\AA~ indicated.  In both cases, the absence of significant molecular and dust opacities allows the molecular lines and resonance line to be much stronger than in the standard and dusty models.  The JOLA break in the cleared-dust models is also much stronger than in the standard and dusty models. } 

\plotone{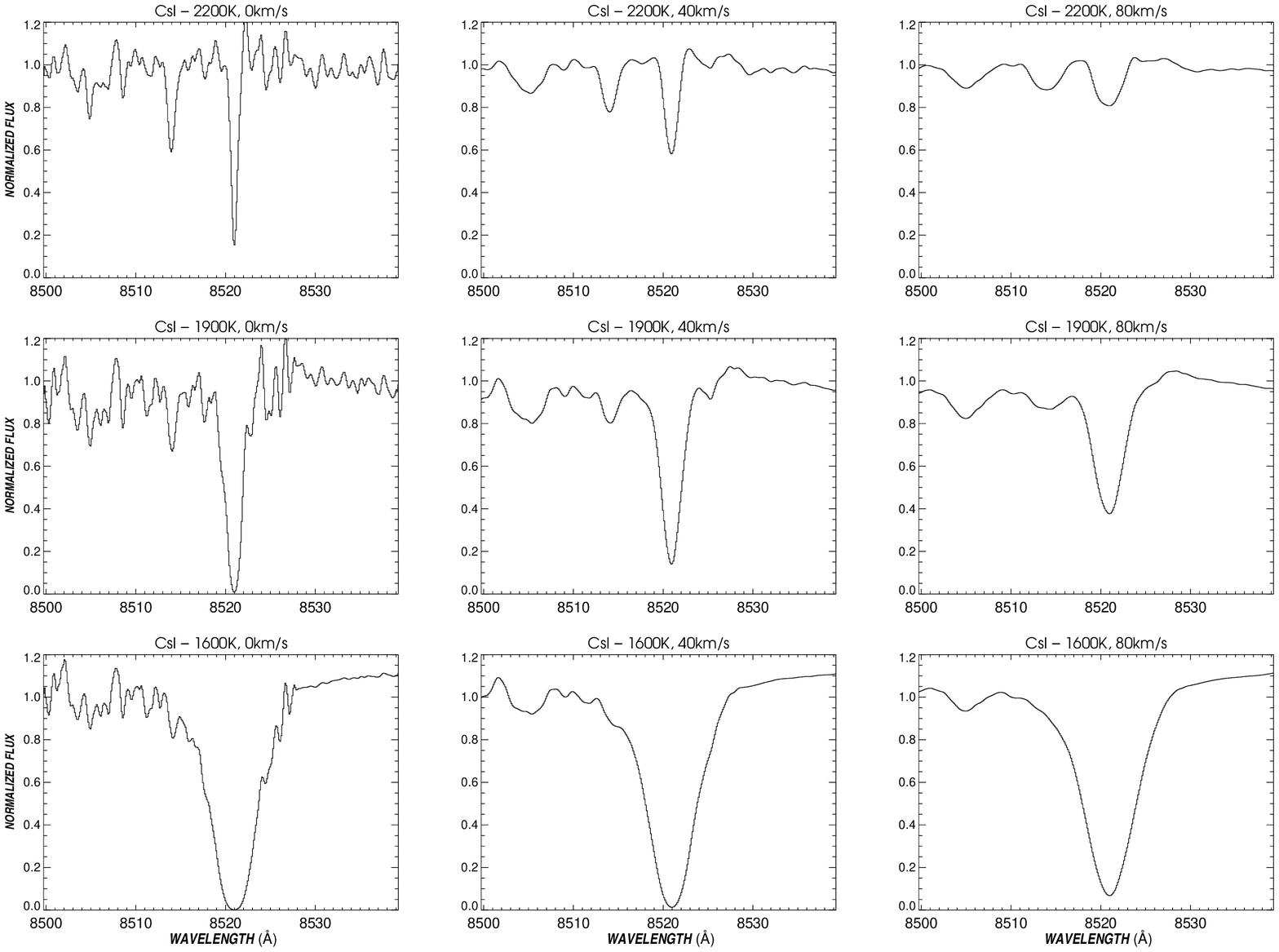}
\figcaption{\label{tvsini} Effect of \teff and \vsini on line profile, for cleared-dust models. Temperatures of 1600, 1900 and 2200K are shown from bottom to top; \vsini of 0, 40 and 80 \kms are shown from left to right.  Decreasing \teff causes the \cs line to become stronger, increasing in both depth and width.  Increasing \vsini also increases the width of the line, but decreases its depth, thus making it possible to distinguish between changes in \teff and changes in \vsini.  At \teff $\lesssim$ than 1600K (last row), collisional broadening of the resonance line overwhelms the rotational broadening, and it is no longer possible to infer \vsini from the resonance line profile.  The sharpness of molecular lines, on the other hand, is relatively unaffected by \teff, but highly influenced by \vsini; thus molecular lines can be used to infer \vsini information even at low \teff.   } 

\plotone{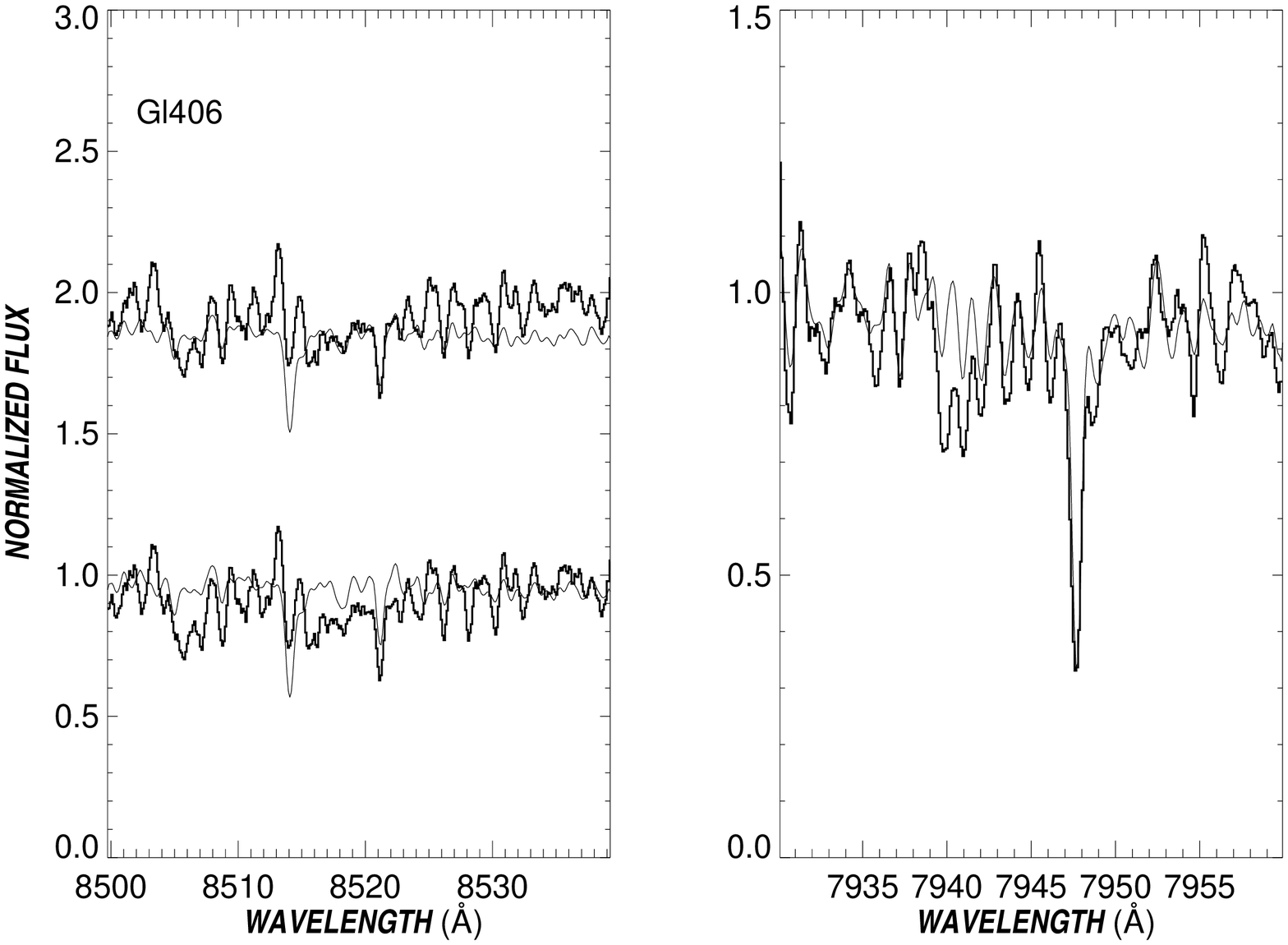}
\figcaption{\label{gl406} {\it Left panel}: Bottom - Gl 406 \cs spectrum (thick solid histogram), with 2800K (thin solid line) cleared-dust model at 0 \kms.  Top - Same, but with the model now renormalized to the spectral interval 8516-8524A~, to give a better fit to the observed line.  The observed spectrum and the model (after renormalization) are offset by 1 from the bottom panel.  {\it Right panel}: Gl 406 \rb spectrum (thick solid hist.), with 2700K cleared-dust model at 0 \kms (thin solid line). } 

\plotone{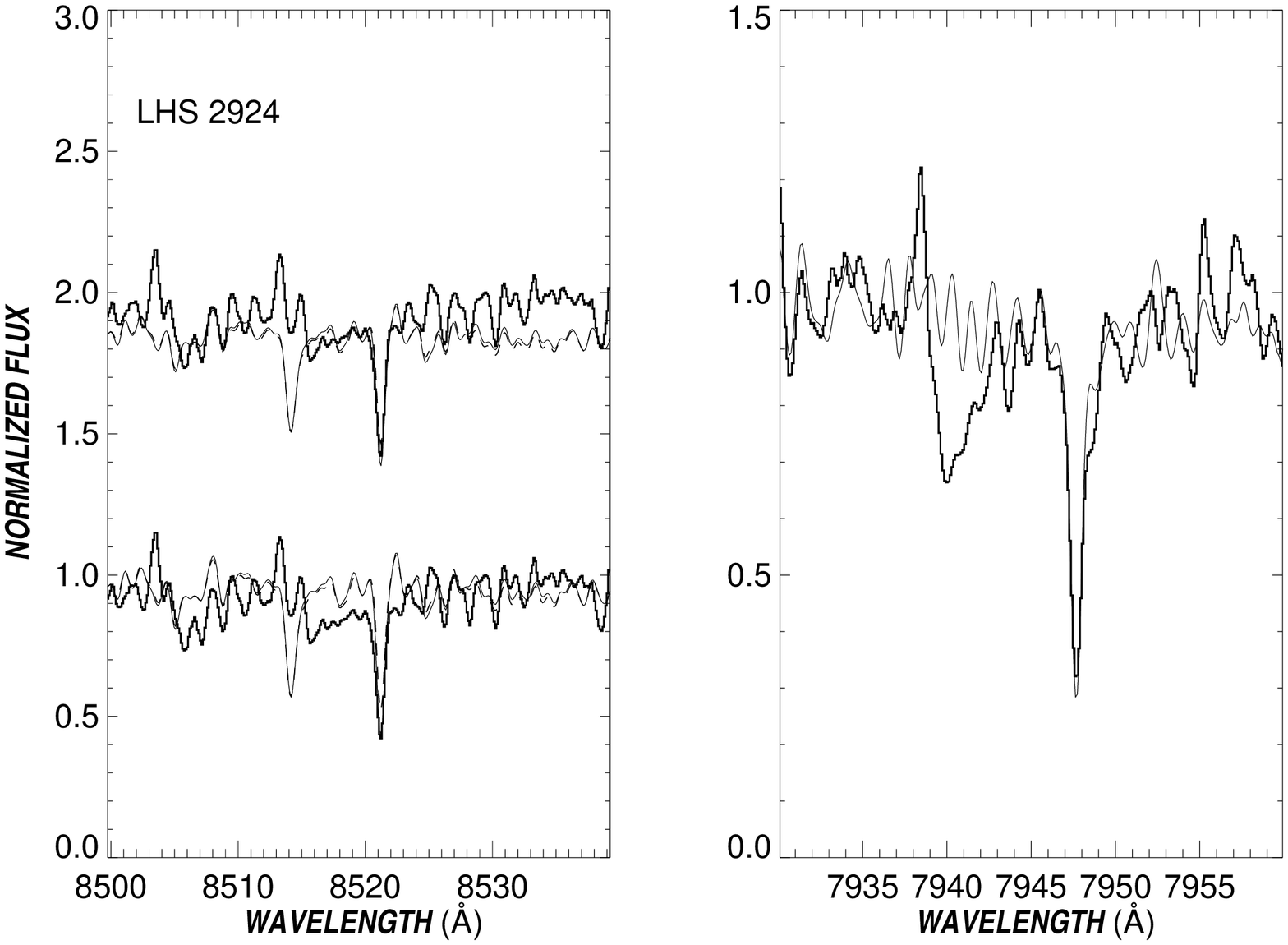}
\figcaption{\label{lhs2924}{\it Left panel}: LHS 2924 \cs spectrum, with 2400K (thin solid line) and 2500K (thin dashed line) cleared-dust models at 10 \kms. {\it Right panel}: LHS 2924 \rb spectrum, with 2600K cleared-dust model at 10 \kms. }

\plotone{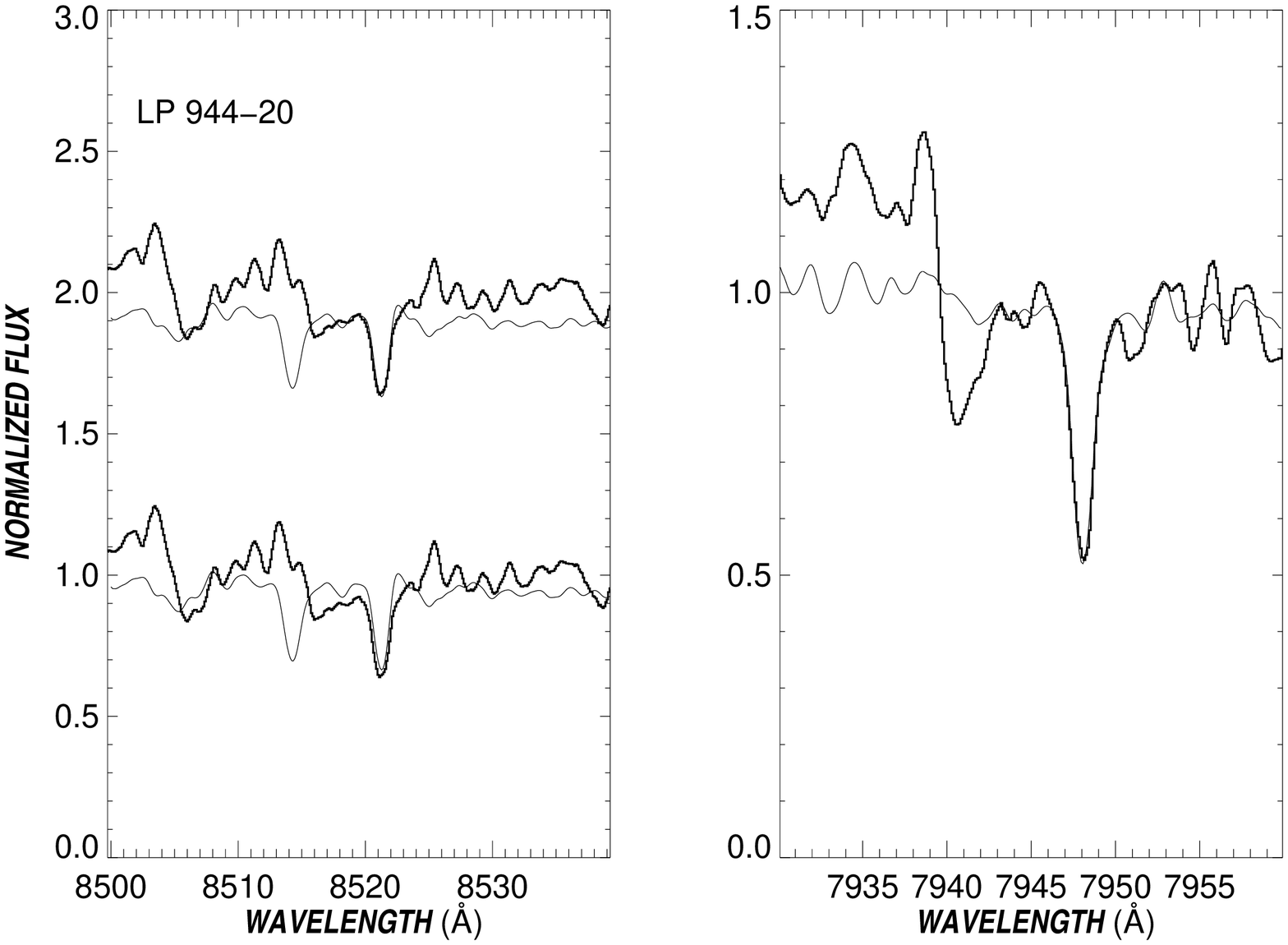}
\figcaption{\label{lp944}{\it Left panel}: LP 944-20 \cs spectrum, with 2400K cleared-dust model at 30 \kms. {\it Right panel}: LP 944-20 \rb spectrum,  with 2600K cleared-dust model at 30 \kms. }

\plotone{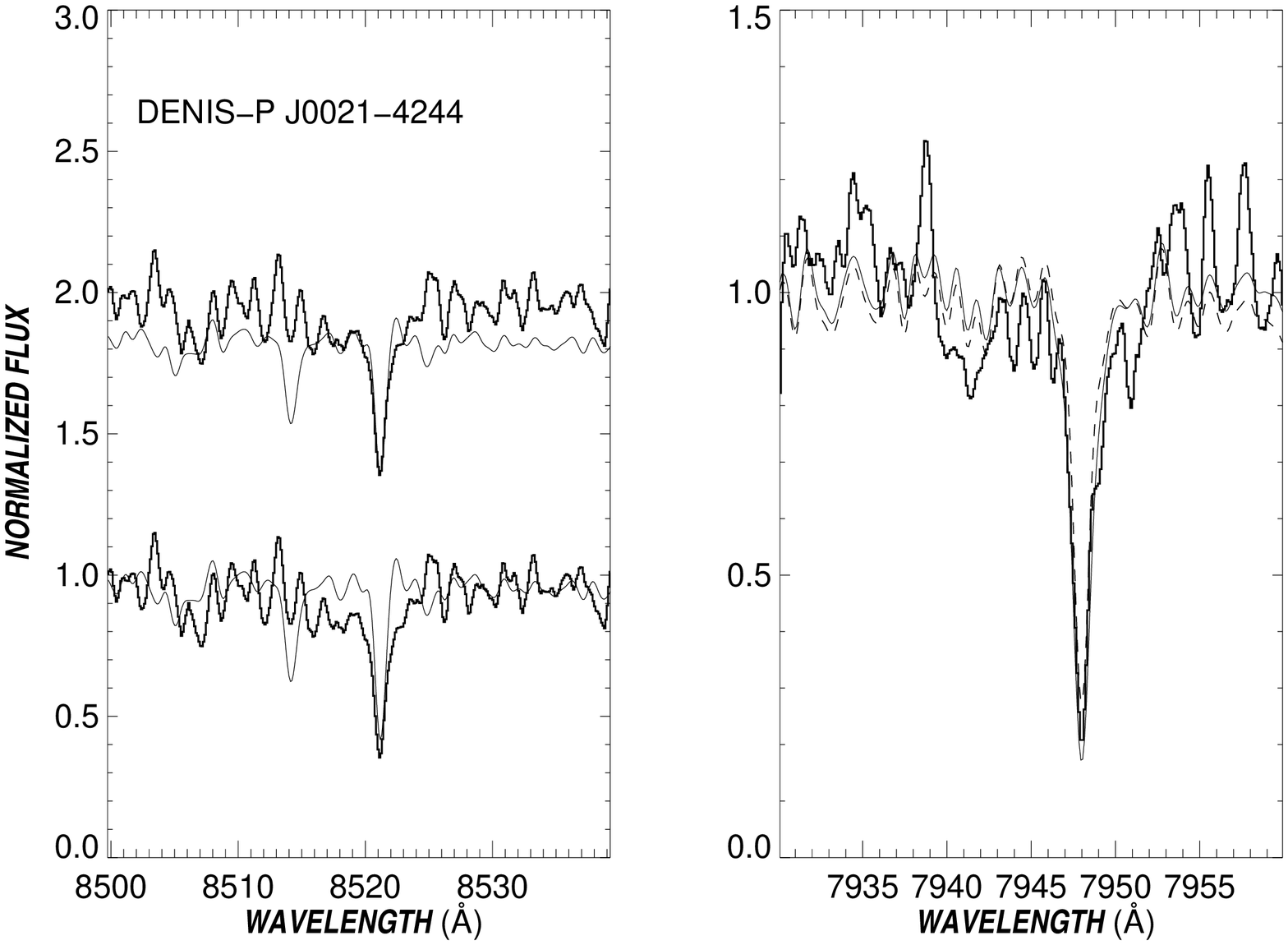}
\figcaption{\label{d0021}{\it Left panel}: DENIS-P J0021-4244 \cs spectrum, with 2300K cleared-dust model at 17.5 \kms. {\it Right panel}: DENIS-P J0021-4244 \rb spectrum, with 2400K (thin solid line) and 2500K (thin dashed line) cleared-dust models at 17.5 \kms. }

\plotone{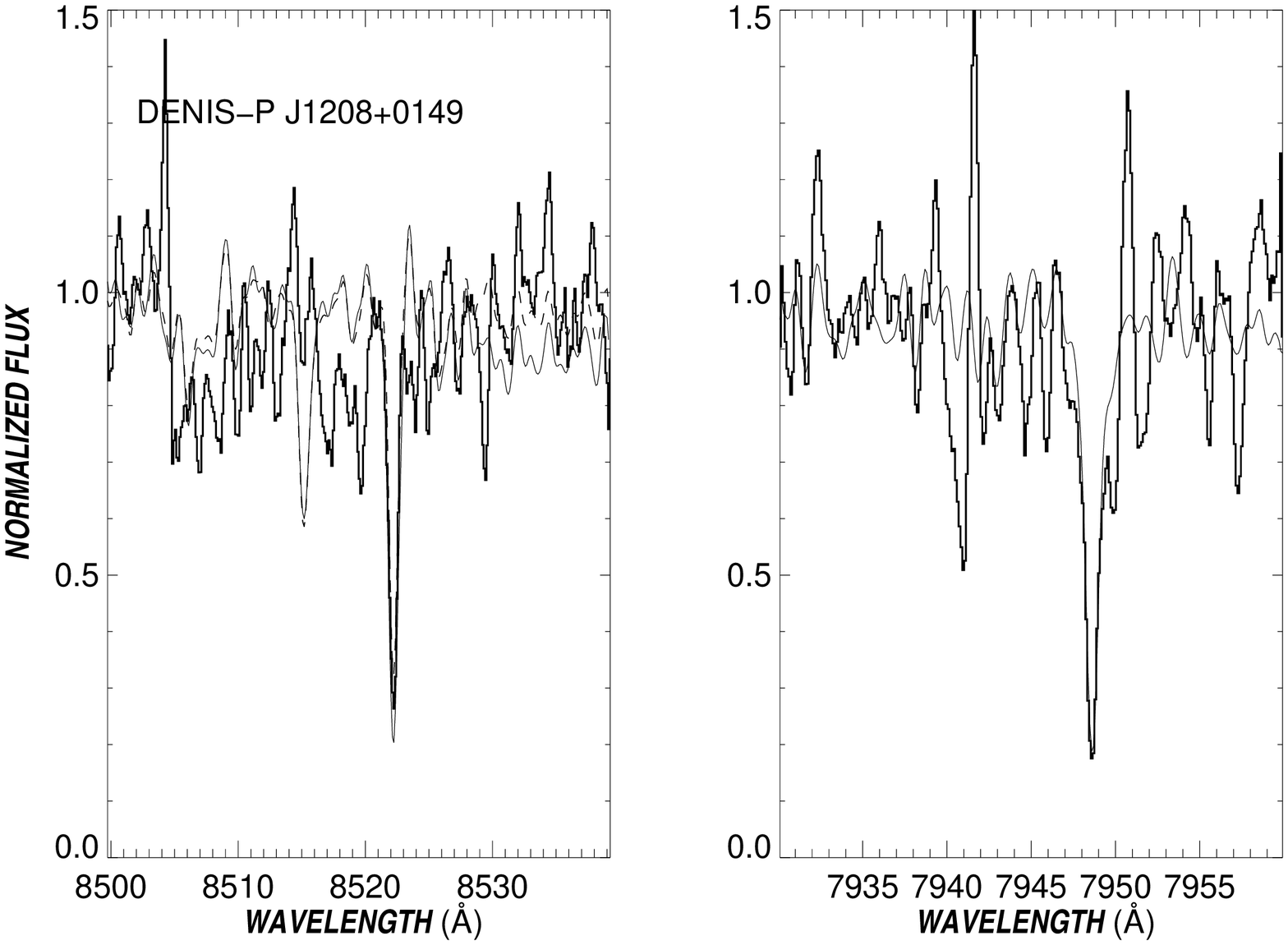}
\figcaption{\label{den1208} {\it Left panel}: DENIS-P J1208+0149 \cs spectrum, with 2200K (thin solid line) and 2300K (thin dashed line) cleared-dust models at 10 \kms.  {\it Right panel}: DENIS-P J1208+0149 \rb spectrum, with 2500K cleared-dust model at 10 \kms.}

\plotone{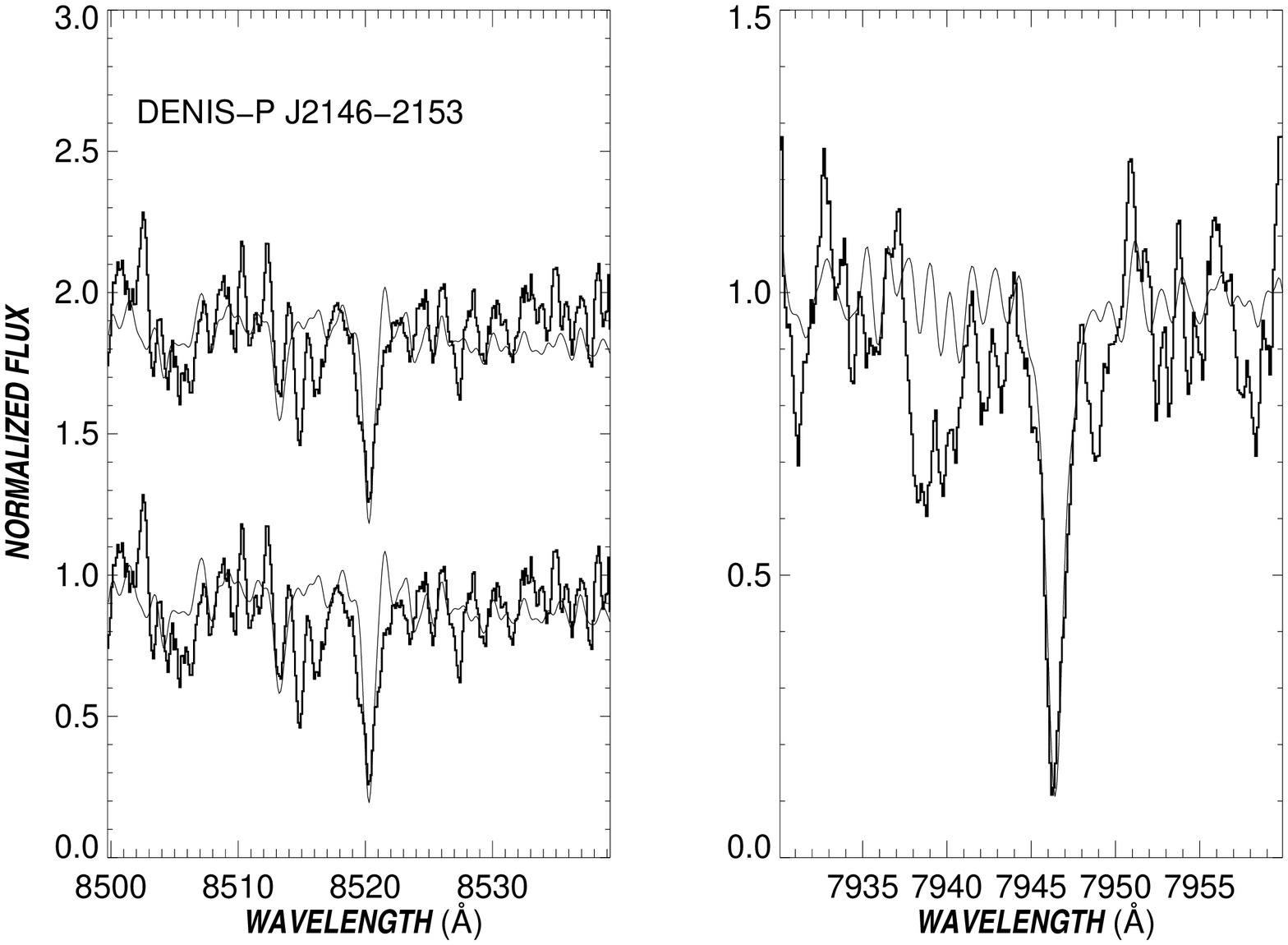}
\figcaption{\label{d2146}{\it Left panel}: DENIS-P J2146-2153 \cs spectrum , with 2200K cleared-dust model at 10 \kms.  {\it Right panel}: DENIS-P J2146-2153 \rb spectrum, with 2400K cleared-dust model at 10 \kms. }

\plotone{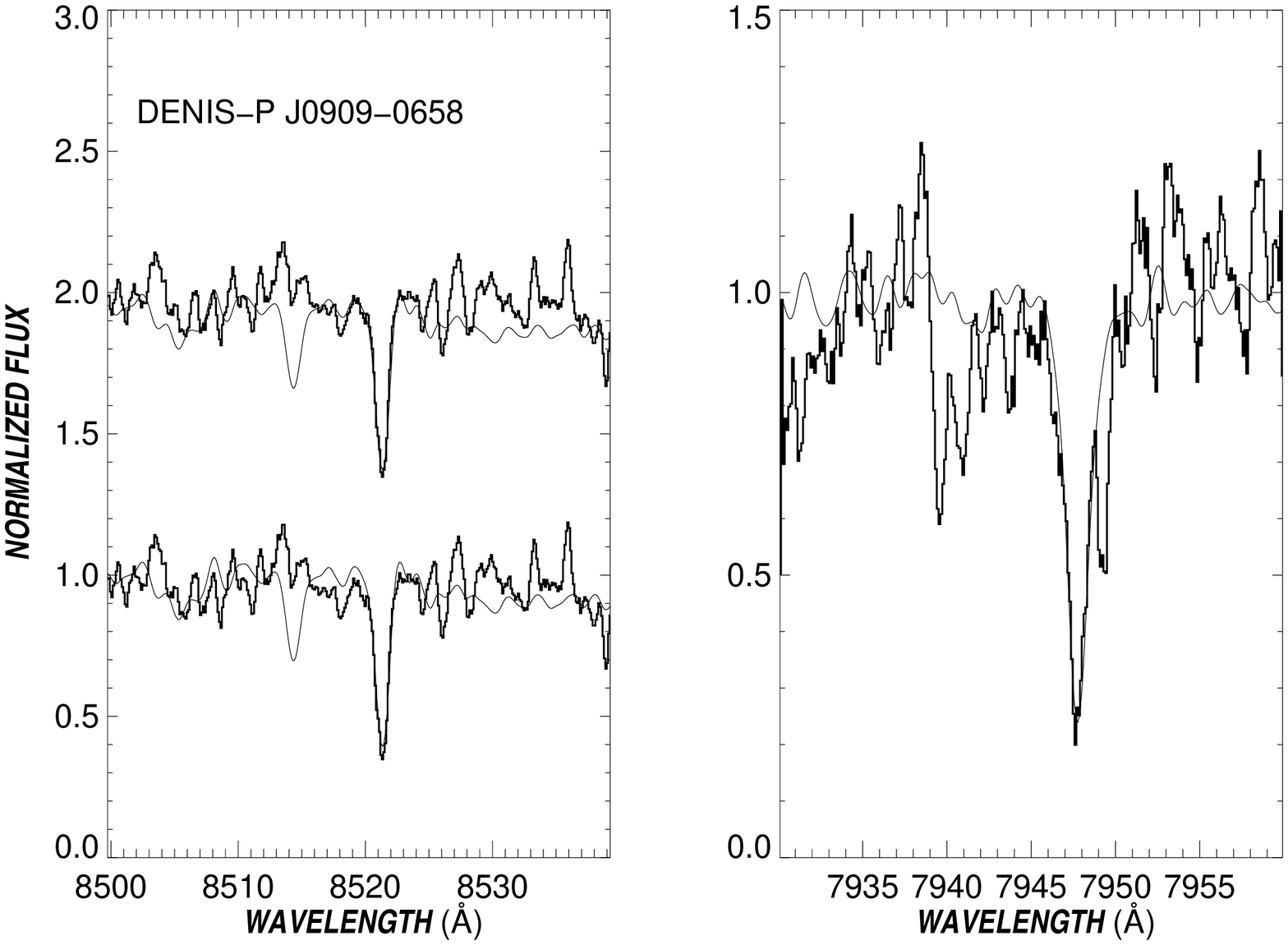}
\figcaption{\label{d0909} {\it Left panel}: DENIS-P J0909-0658 \cs spectrum, with 2200K cleared-dust model at 25 \kms.  {\it Right panel}: DENIS-P J0909-0658 \rb spectrum, with 2400K cleared-dust models at 25 \kms. }

\plotone{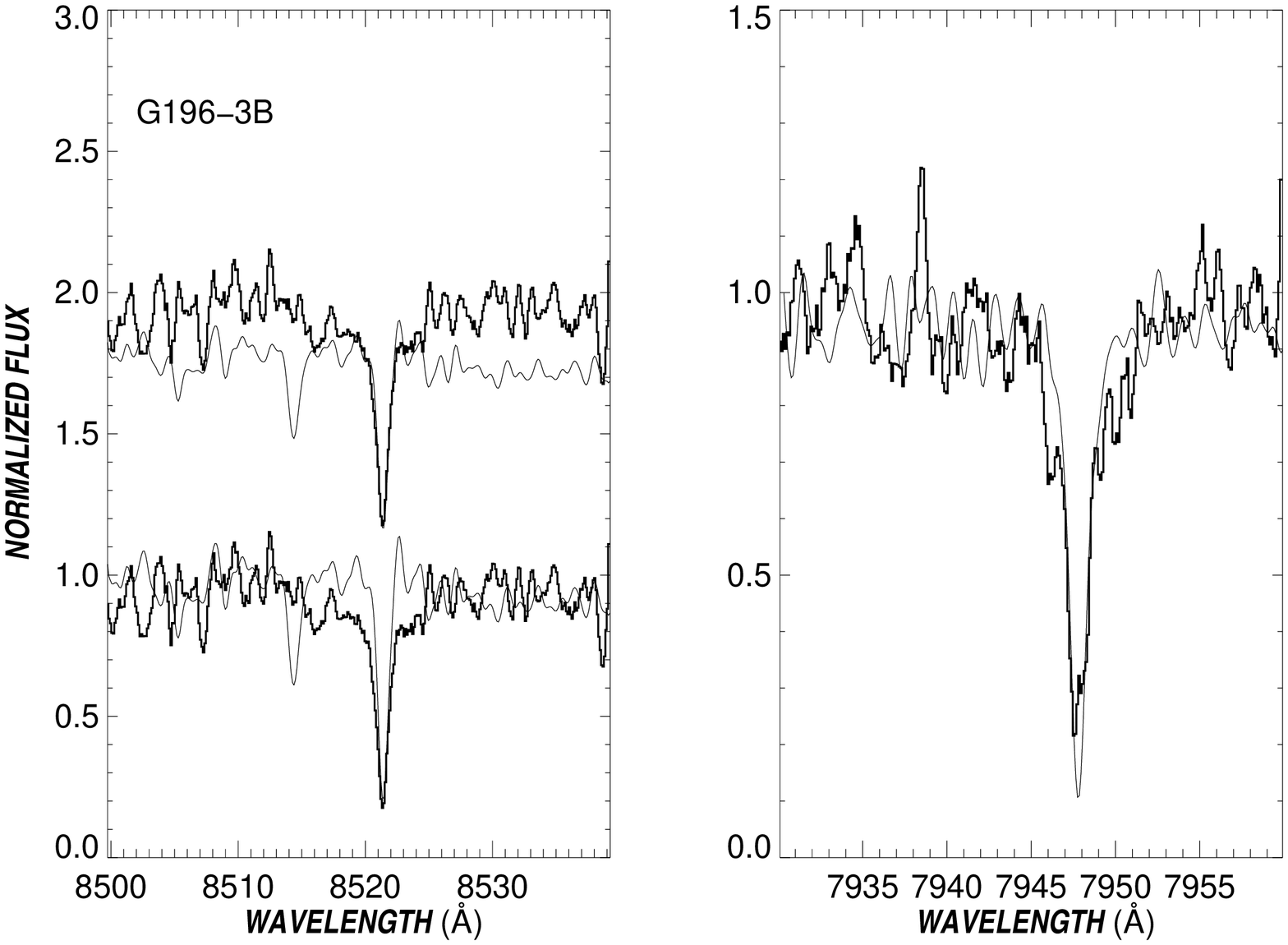}
\figcaption{\label{g1963b} {\it Left panel}: G 196-3B \cs spectrum , with 2200K cleared-dust model at 10 \kms.  {\it Right panel}: G 196-3B \rb spectrum, with 2400K cleared-dust model at 10 \kms.  }

\plotone{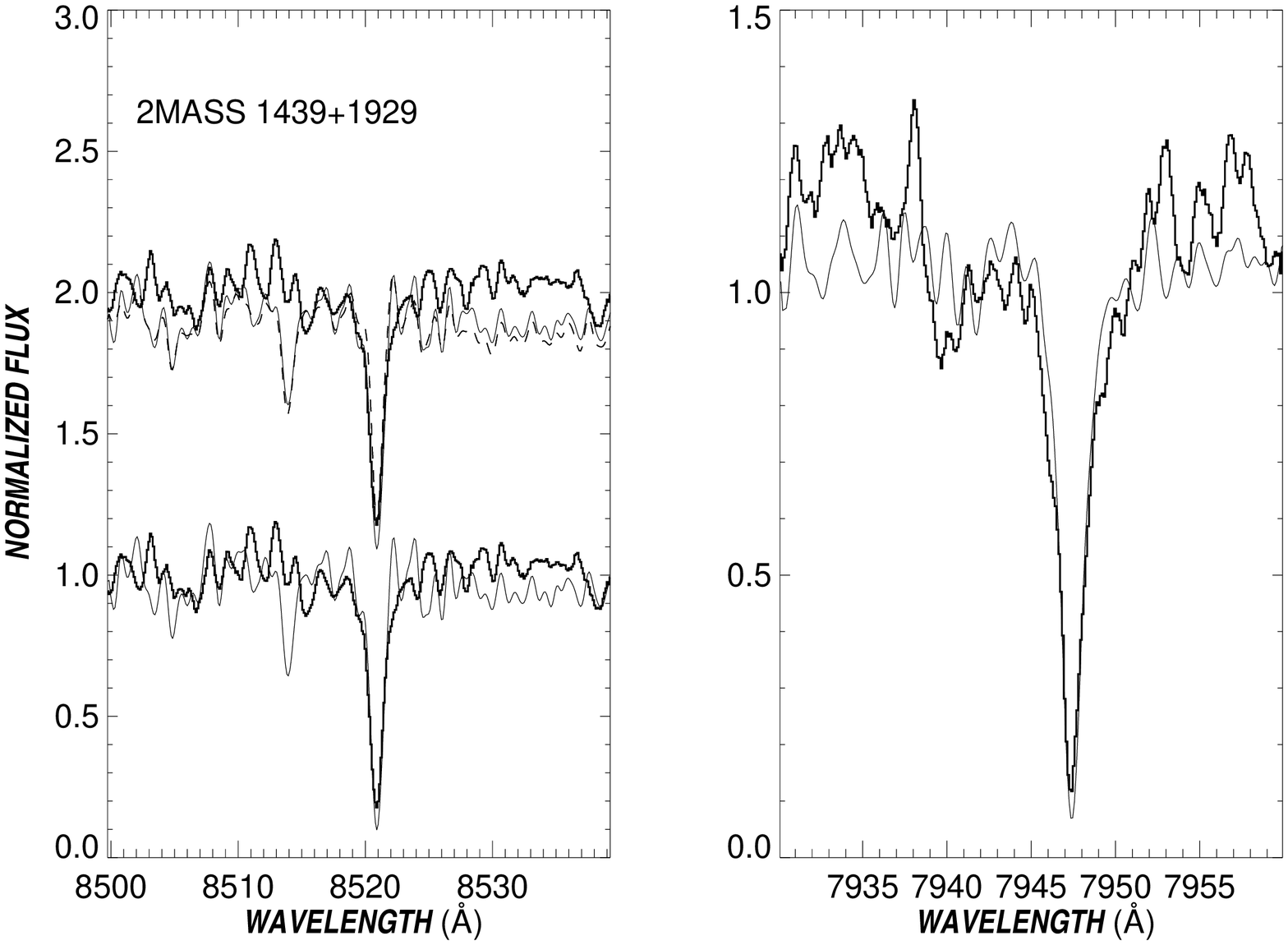}
\figcaption{\label{mass1439} {\it Left panel}: 2MASSW 1439+1929 \cs spectrum, with 2100K (thin solid line) and 2200K (thin dashed line) cleared-dust model at 10 \kms.  {\it Right panel}: 2MASSW 1439+1929 \rb spectrum, with 2300K cleared dust model at 10 \kms.  }

\plotone{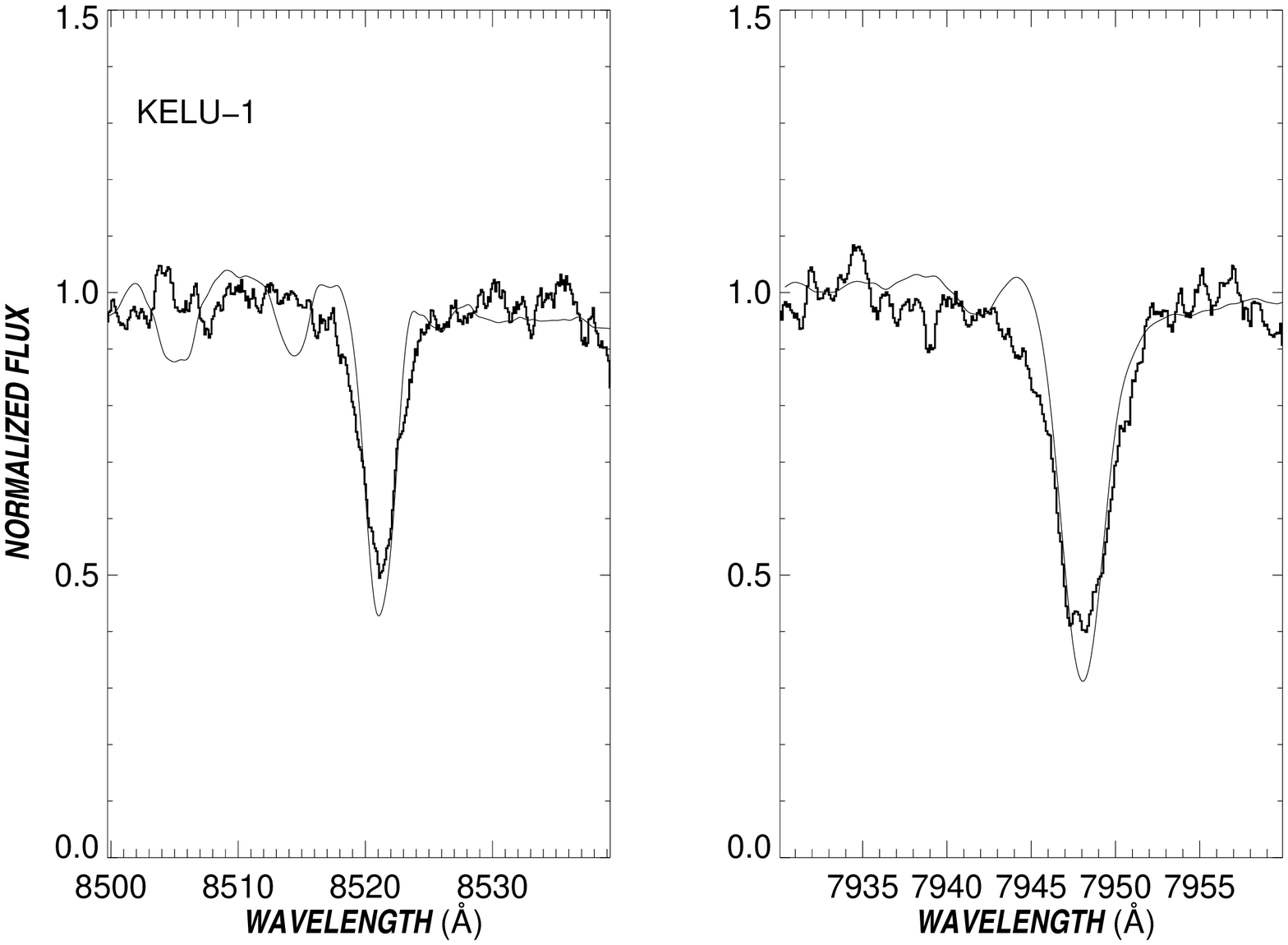}
\figcaption{\label{kelu1} {\it Left panel}: Kelu-1 \cs spectrum , with 2000K (thin solid line) cleared-dust model at 60 \kms.  {\it Right panel}: Kelu-1 \rb spectrum, with 2200K (thin solid line) cleared-dust model at 60 \kms.  This object is the fastest rotator in our sample. }

\plotone{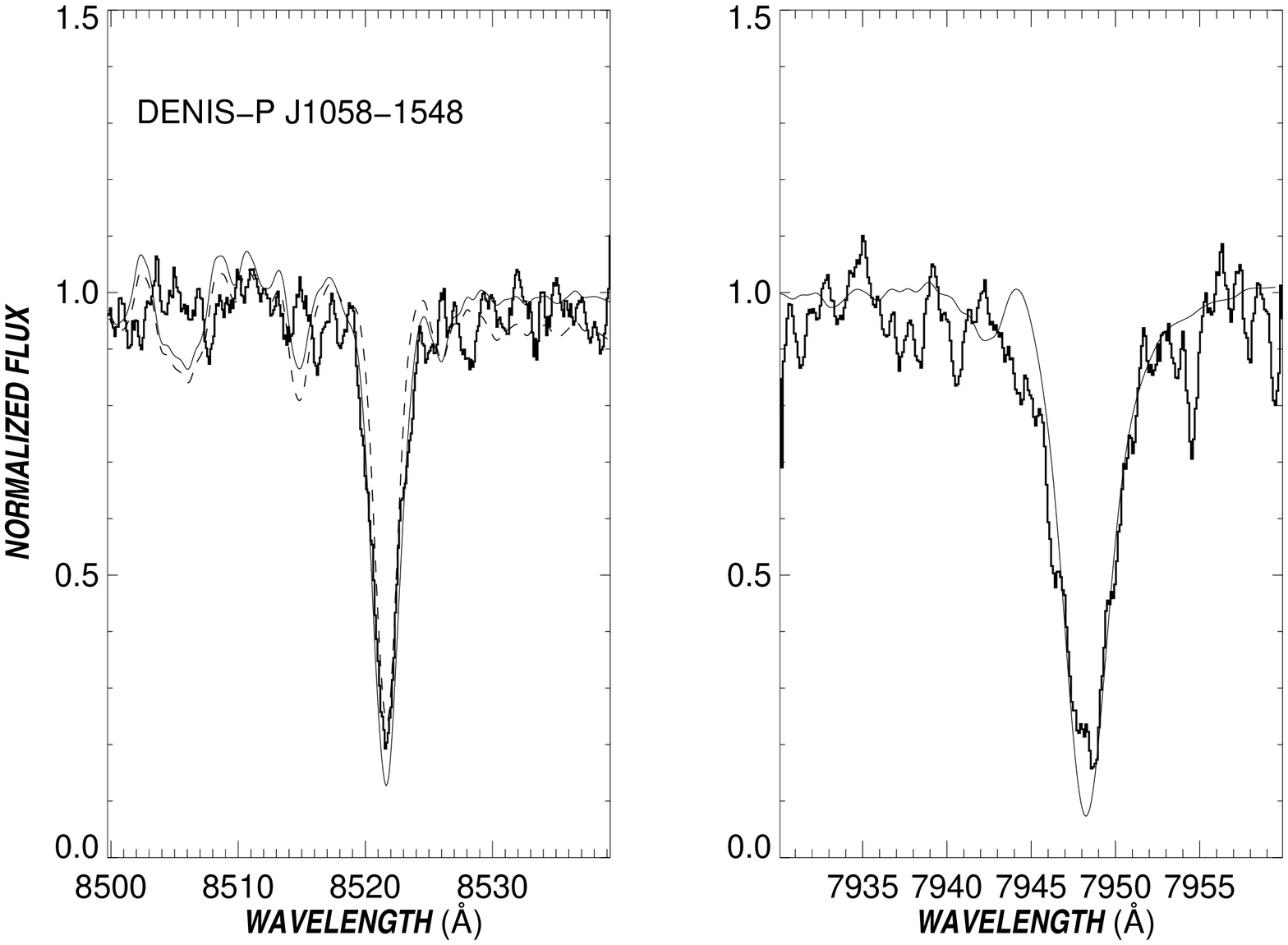}
\figcaption{\label{d1058} {\it Left panel}: DENIS-P J1058-1548 \cs spectrum, with 1900K (thin solid line) and 2000K (thin dashed line) cleared-dust models at 37.5 \kms.  {\it Right panel}: DENIS-P J1058-1548 \rb spectrum, with 2000K cleared-dust model at 37.5 \kms.  }

\plotone{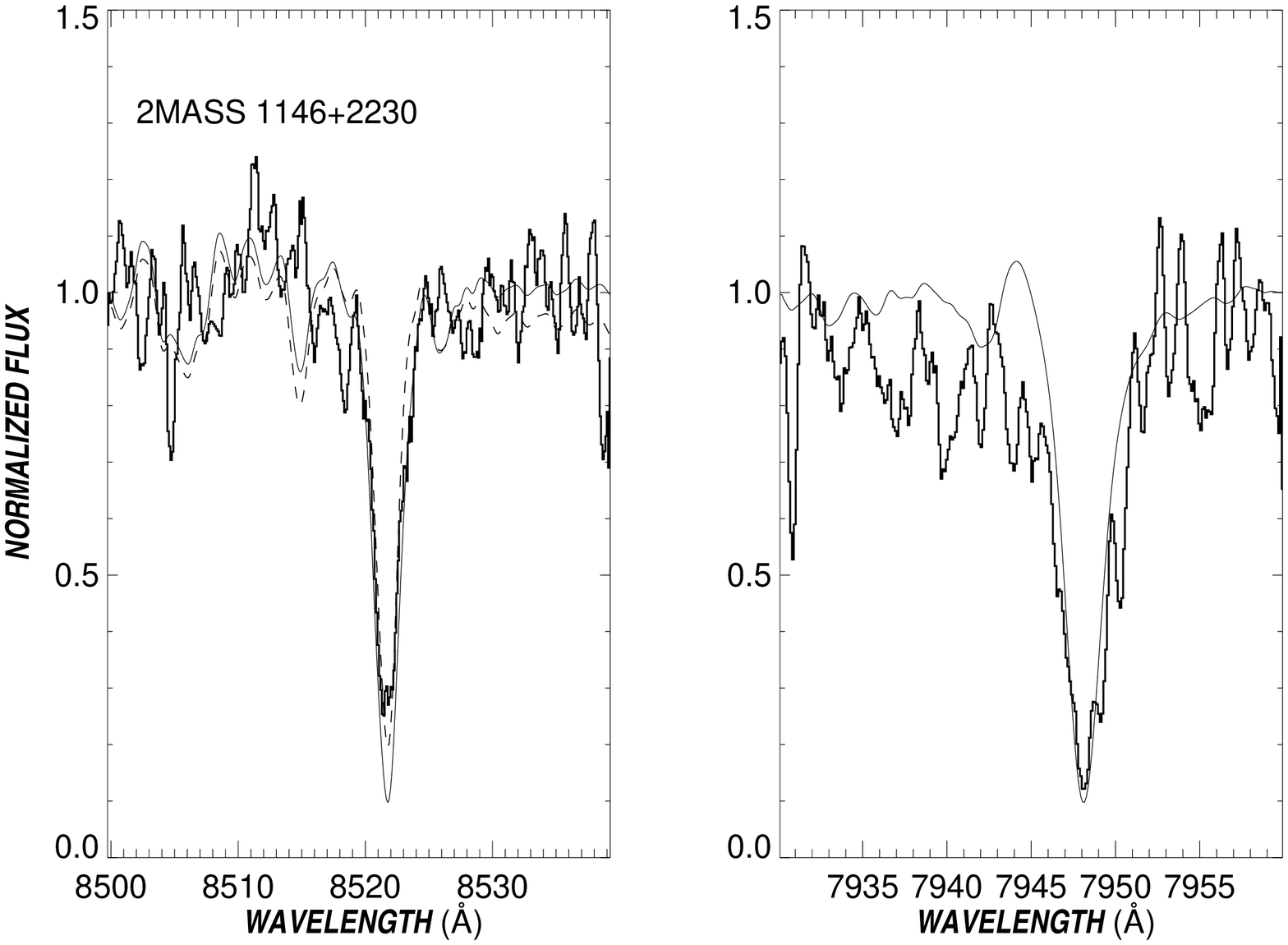}
\figcaption{\label{mass1146} {\it Left panel}: 2MASSW 1146+2230 \cs spectrum, with 1900K (thin solid line) and 2000K (thin dashed line) cleared-dust model at 32.5 \kms.  {\it Right panel}: 2MASSW 1146+2230 \rb spectrum, with 2100K cleared dust model at 32.5 \kms.  }

\plotone{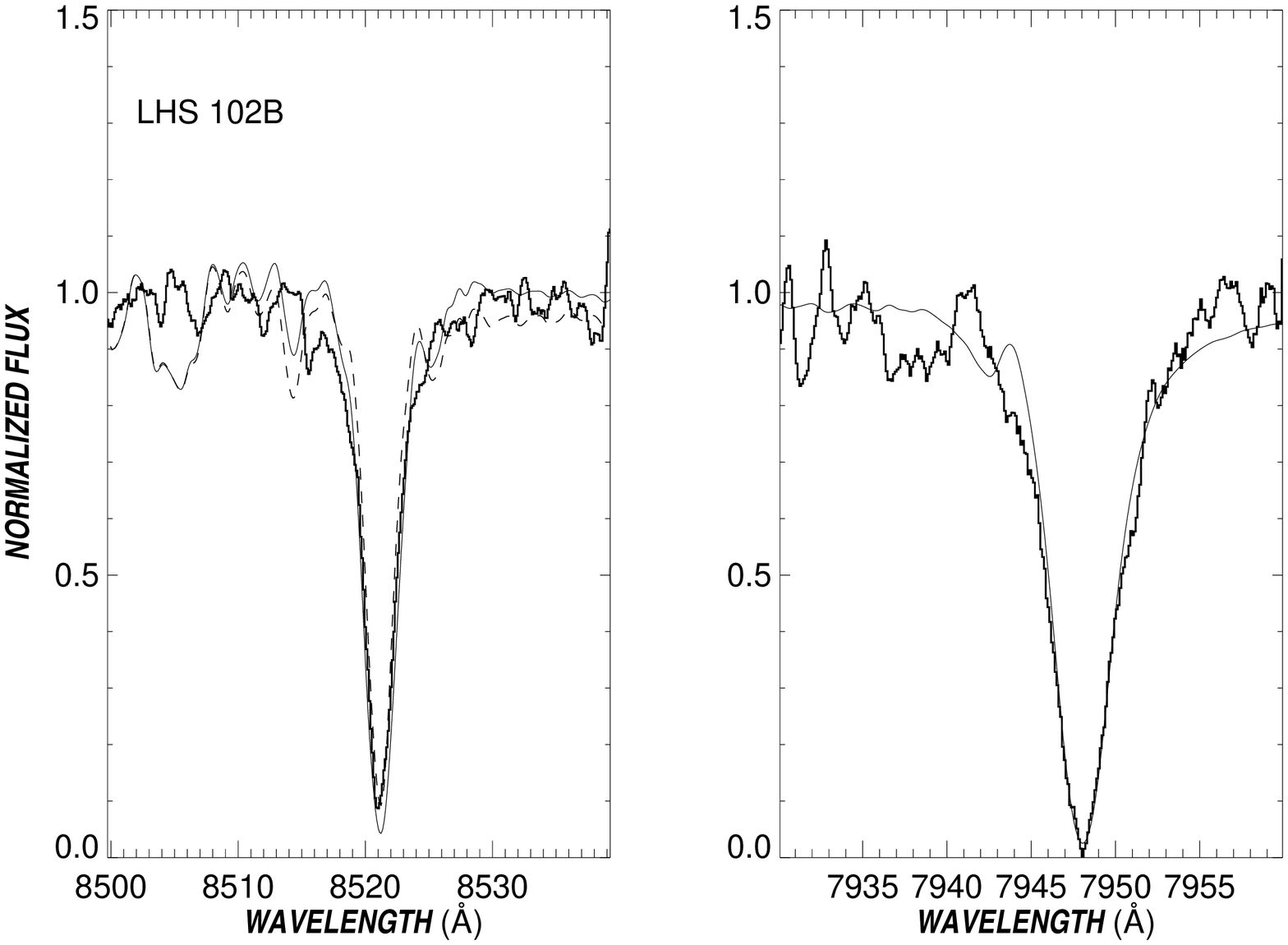}
\figcaption{\label{lhs102b} {\it Left panel}: LHS 102B \cs spectrum with 1800K (thin solid line) and 1900K (thin dashed line) cleared-dust models at 32.5 \kms.  {\it Right panel}: LHS 102B \rb spectrum, with 1900K cleared-dust model at 
32.5 \kms. }

\plotone{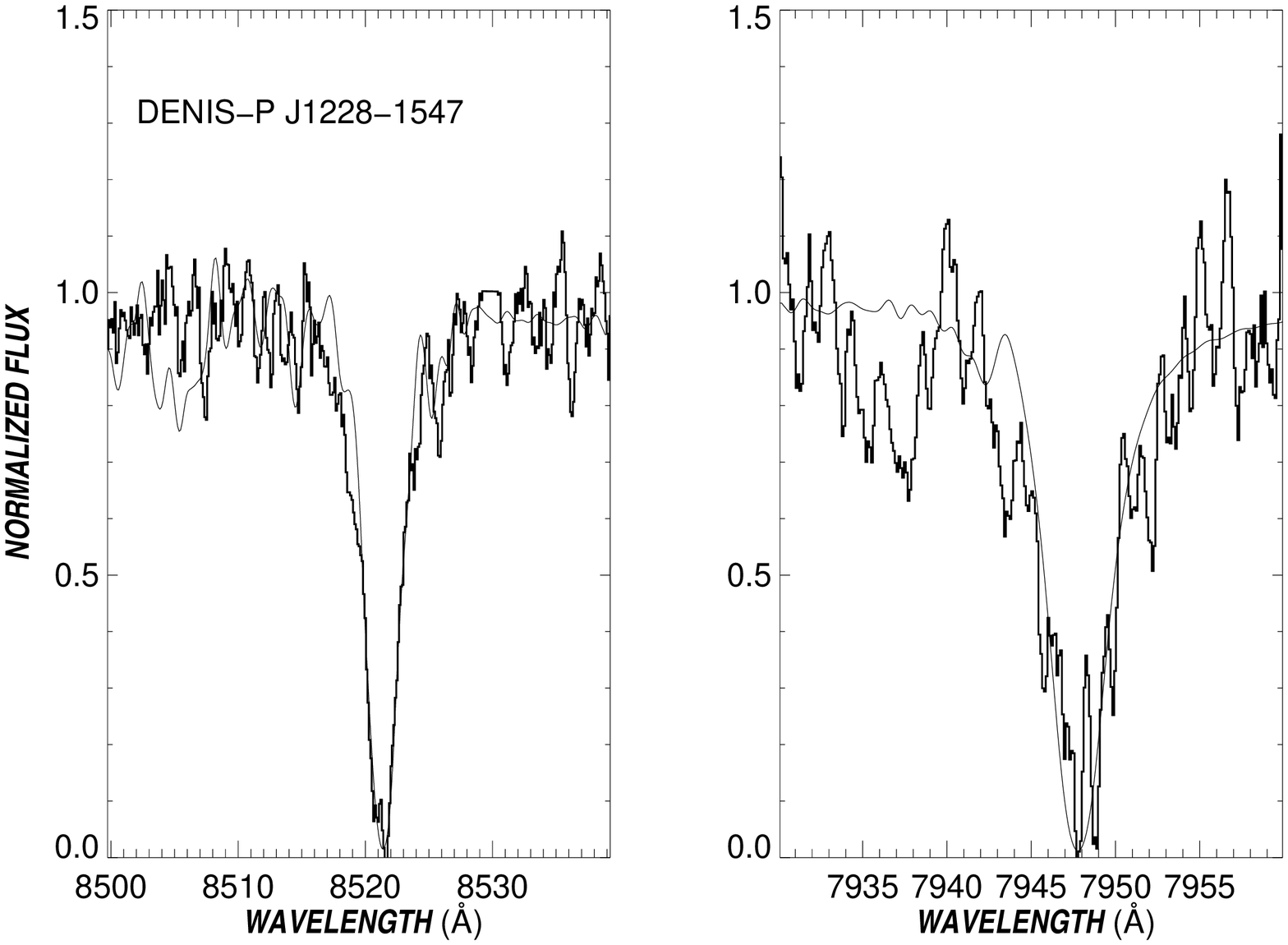}
\figcaption{\label{den19} {\it Left panel}: DENIS-P J1228-1547 \cs spectrum , with 1800K cleared-dust model at 22 \kms.  {\it Right panel}: DENIS-P J1228-1547 \rb spectrum, with 1900K cleared-dust models at 22 \kms.  Both spectra appear noisy, and the \rb fit is crude.  This object is a binary, and it is possible that we are looking at a composite spectrum. }

\plotone{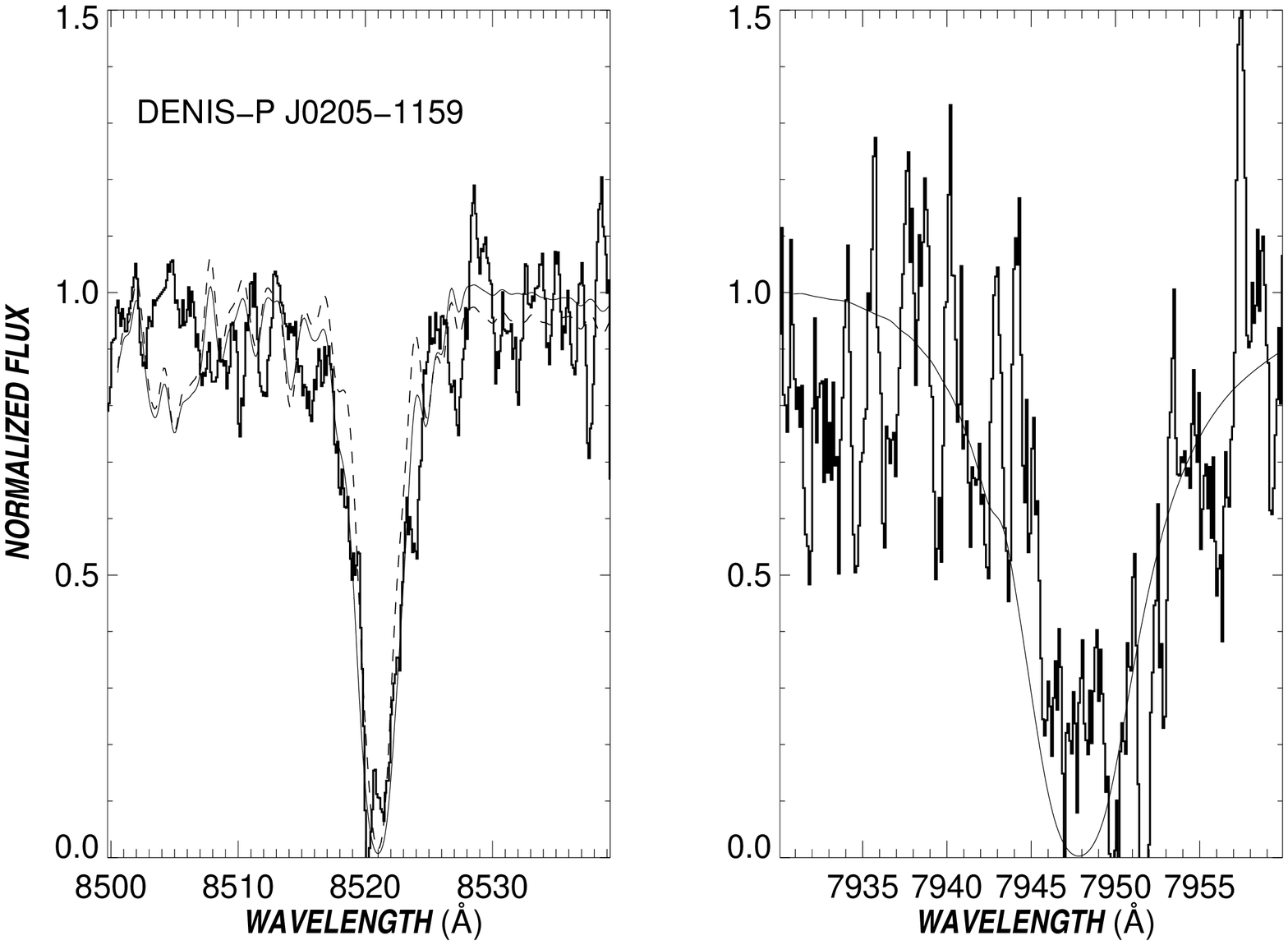}
\figcaption{\label{d0205} {\it Left panel}: DENIS-P J0205-1159 \cs spectrum , with 1700K (thin solid line) and 1800K (thin dashed line) cleared-dust models at 22 \kms.  {\it Right panel}: DENIS-P J0205-1159 \rb spectrum, with 1700K cleared-dust model at 22 \kms.  Due to the low S/N of the \rb spectrum, no best fit was found; the model shown is for comparison only to the \cs fit at the same \teff and rotational velocity. } 

\plotone{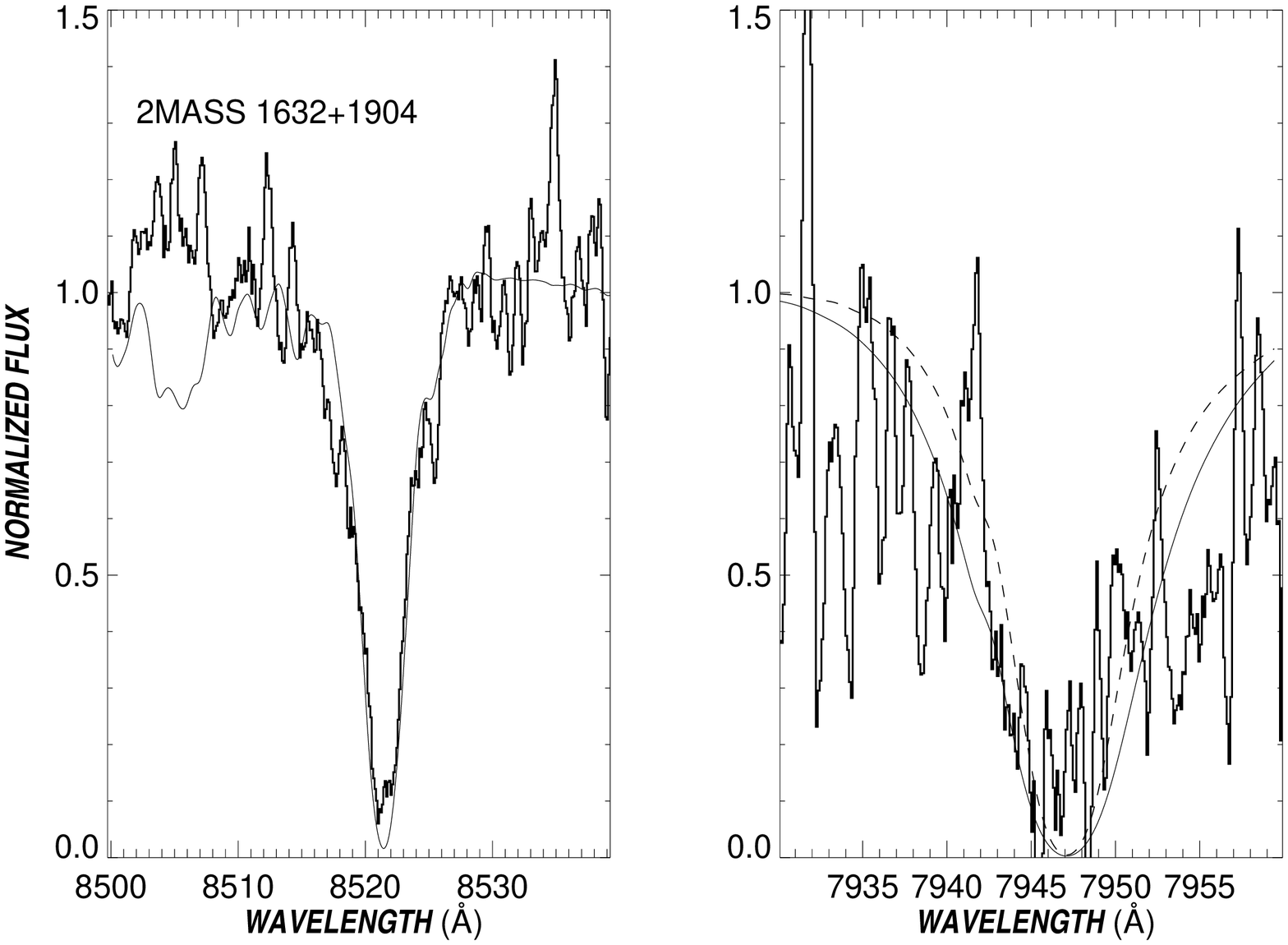}
\figcaption{\label{mass1632} {\it Left panel}: 2MASSW 1632+1904 \cs spectrum, with 1700K cleared-dust model at 30 \kms.  {\it Right panel}: 2MASSW 1632+1904 \rb spectrum, with 1600K (thin solid line) and 1700K (thin dashed line) cleared dust models at 30 \kms.  Due to the low S/N of the \rb spectrum, no best fit was found; the models are shown as only very tentative fits, and for comparison to the \cs fit.  }

\plotone{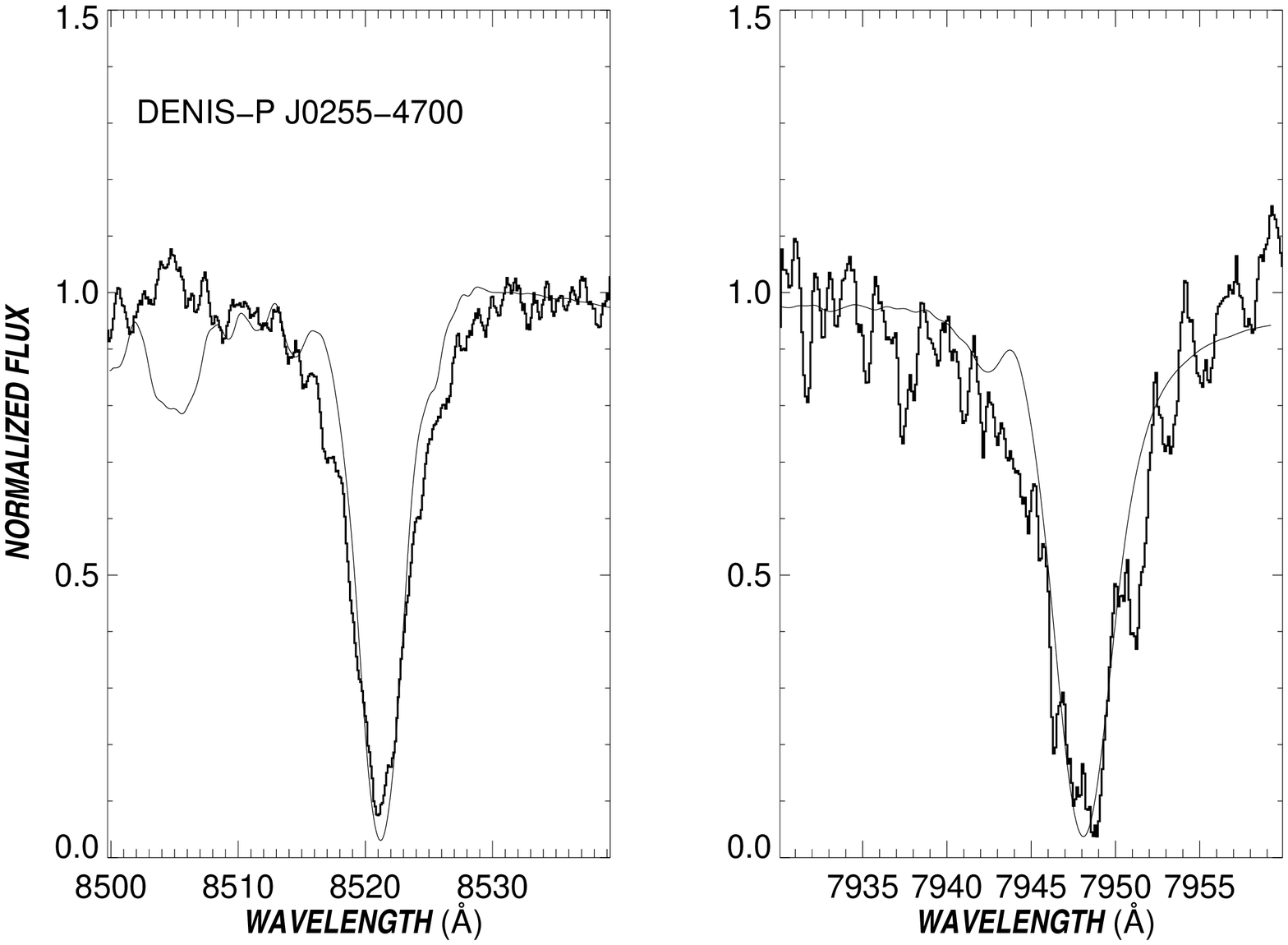}
\figcaption{\label{d0255} {\it Left panel}: DENIS-P J0255-4700 \cs spectrum , with 1700K (thin solid line) cleared-dust model at 40 \kms.  {\it Right panel}: DENIS-P J0255-4700 \rb spectrum, with 1900K cleared-dust model at 40 \kms.  }

\clearpage

\begin{deluxetable}{lccccccc}
\scriptsize
\tablecaption{\label{tab1} Observational Sample.}
\tablewidth{0pt}
\tablehead{
\colhead{object}  &  
\colhead{obs. date} &
\colhead{spectral type}  &
\colhead{I-K}  &
\colhead{$v_{rad}$}  \nl 
 & & & & km s$^{-1}$ \nl}   
                            
\startdata
Gl 406 (LHS 36) & 7 Dec 1997 & M6 & 3.31\tablenotemark{a} & +19$\pm$1 \nl  & 2 Jun 1997 &  &   & +43$\pm$1 \nl  & 3 Jun 1997 &  &   & +40$\pm$1 \nl 
      & 10 Jun 1999 &  &  & +19$\pm$1 \nl  & 11 Dec 1993 & &  & +18$\pm$1 \nl
LHS 2924 & 2,4 Jun 1997 & M9 & 4.47\tablenotemark{b} & -37$\pm$1 \nl
LP 944-20 & 20 Sep 1998 & bdM9 & 4.63\tablenotemark{b} & +6$\pm$1 \nl
DENIS-P J1208+0149 & 9 Jun 1999 & M9 & 4.61$\pm$0.20\tablenotemark{d} & +10$\pm$1 \nl

DENIS-P J0021-4244 & 20 Sep 1998 & M9.5 & 4.58$\pm$0.08\tablenotemark{c} & +2$\pm$1 \nl
DENIS-P J2146-2153 & 24,25 May 1998 & L0 & -- & -3$\pm$1 \nl  & 9 Jun 1999 & &  & -4$\pm$1 \nl
DENIS-P J0909-0658 & 7 Dec 1997 & L0 & 4.70$\pm$0.12\tablenotemark{c} & +27$\pm$1 \nl
G196-3B & 24,25 May 1998 & bdL1: & 5.79$\pm$0.03\tablenotemark{e} & 0$\pm$1 \nl  & 10 Jun 1999 & &  &  -3$\pm$1\nl
2MASSW 1439+1929 & 10 Jun 1999 & L1 & 4.44\tablenotemark{f} & -29$\pm$1 \nl
Kelu-1 & 2 Jun 1997 & bdL2 & 5.0\tablenotemark{b} & +17$\pm$1 \nl
DENIS-P J1058-1548 & 2 Jun 1997 & L2.5 & 5.19\tablenotemark{b} & +19$\pm$2 \nl
2MASSW 1146+2230 & 10 Jun 1999 & L3 & 4.99\tablenotemark{f} & -16$\pm$1 \nl
LHS 102B & 20 Sep 1998 & L4 & 5.60$\pm$0.10\tablenotemark{g} & +30$\pm$1 \nl
DENIS-P J1228-1547 & 2 Jun 1997 & bdL4.5 & 5.43\tablenotemark{b} & +4$\pm$1 \nl  & 10 Jun 1999 & &  &  +1$\pm$1\nl

DENIS-P J0205-1159 & 7 Dec 1997 & L5 & 5.28\tablenotemark{b} & +6: \nl  & 20 Sep 1998 & &  &  +7$\pm$2\nl
2MASSW 1632+1904 & 9,10 Jun 1999 & bd:L6 & 6.00\tablenotemark{f}& +5: \nl
DENIS-P J0255-4700 & 20 Sep 1998 & bd:L6 & 5.28$\pm$0.10\tablenotemark{g} & +13$\pm$3 \nl
\enddata
\tablenotetext{a}{\cite{Basri95}}
\tablenotetext{b}{\cite{Leggett98}}
\tablenotetext{c}{\cite{Delfosse97}}
\tablenotetext{d}{\cite{Martin99b}}
\tablenotetext{e}{\cite{Rebolo98}}
\tablenotetext{f}{\cite{Kirkpatrick99a}; this is I$_{spec}$-K$_s$}
\tablenotetext{g}{Delfosse, private communication}
\end{deluxetable}

\clearpage 
 
\begin{deluxetable}{lcccccccc}
\scriptsize
\tablecaption{\label{tab2} Derived Characteristics.}
\tablewidth{0pt}
\tablehead{
\colhead{Object}  &  
\colhead{{\it v}~sin{\it i}}  & 
\colhead{Spec. T.}  &
\colhead{\teff}  &
\colhead{\teff}  &
\colhead{\teff}  &
\colhead{EqW}  &
\colhead{EqW}  \nl
 &  & & \cs & \rb & adopted & \cs & \rb \nl
 & km s$^{-1}$ & & K & K & K & A~ & A~ \nl}   
                            
\startdata
Gl 406 & $<$3 & M6\tablenotemark{a} & 2800 & 2700 & 2800 & 0.16 & 0.56-0.67: \nl
LHS 2924 & 10$\pm$1 & M9\tablenotemark{a} & 2400-2500 & 2600 & 2450\tablenotemark{b} & 0.43 & 0.75-0.92: \nl
LP 944-20 & 30$\pm$2.5 & bdM9\tablenotemark{c} & 2400 & 2600 & 2400 & 0.6& 0.94 \nl
DENIS-P J1208+0149 & 10$\pm$2.5 & M9\tablenotemark{a} & 2200-2300 & 2500 & 2250 & 0.65 & 1.2: \nl
DENIS-P J0021-4244 & 17.5$\pm$2.5 & M9.5\tablenotemark{d} & 2300 & 2400-2500 & 2300 & 0.86-0.93: & 1.14: \nl
DENIS-P J2146-2153 & 10$\pm$2.5 & L0\tablenotemark{d} & 2200 & 2400 & 2200 & 1.2 & 1.42-1.58: \nl
DENIS-P J0909-0658 & 25$\pm$2.5 & L0\tablenotemark{a} & 2200 & 2400 & 2200 & 0.85 & 1.5 \nl
G196-3B         & 10$\pm$2.5 & bdL1:\tablenotemark{a} & 2200 & 2400 & 2200: & 1.02 & 1.52-1.84: \nl
2MASSW 1439+1929 & 10$\pm$2.5 & L1\tablenotemark{c} & 2100-2200 & 2300 & 2150 & 1.33 & 2.05 \nl
Kelu-1          & 60$\pm$5 & bdL2\tablenotemark{a} & 2000 & 2200 & 2000 & 1.7 & 2.54 \nl
DENIS-P J1058-1548 & 37.5$\pm$2.5 & L2.5\tablenotemark{a} & 1900-2000 & 2000 & 1950\tablenotemark{b} & 2.2 & 3.35 \nl
2MASSW 1146+2230 & 32.5$\pm$2.5 & L3\tablenotemark{c} & 1900-2000 & 2100 & 1950 & 2.09 & 3.3-3.6: \nl
LHS 102B        & 32.5$\pm$2.5 & L4\tablenotemark{a} & 1800-1900 & 1900 & 1850\tablenotemark{b} & 3.05 & 5.5 \nl
DENIS-P J1228-1547 & 22$\pm$2.5 & bdL4.5\tablenotemark{a} & 1800 & 1900 & 1800 & 3.7 & - \nl
DENIS-P J0205-1159 & 22$\pm$5 & L5\tablenotemark{a} & 1700-1800 & - & 1750\tablenotemark{b} & 4.2 & - \nl
2MASSW 1632+1904 & 30$\pm$10 & bd:L6\tablenotemark{a} & 1700 & - & 1700 & 5.01 & - \nl
DENIS-P J0255-4700 & 40$\pm$10 & bd:L6\tablenotemark{a} & 1700 & 1900 & 1700\tablenotemark{b} & 5.75 & 5.3:: \nl
\enddata
\tablenotetext{a}{\cite{Martin99b}}
\tablenotetext{b}{\cite{Martin99b} attribute to us a slightly different \teff for this object; they used our preliminary fits, which have been revised.}
\tablenotetext{c}{\cite{Kirkpatrick99a}}
\tablenotetext{d}{assigned in this paper, based on our derived \teff and the spectral-type assigned by \cite{Martin99b} to an objects with similar \teff}

\end{deluxetable}


\clearpage

\begin{thebibliography}{}

\bibitem[Allard et al. 1996]{Allard96}
Allard, F., Hauschildt, P.H., Baraffe, I., Chabrier, G., 1996, \apj Letters, 465, L123
\bibitem[Allard et al. 1997]{Allard97}
Allard, F., Hauschildt, P., Alexander, D.R., Starrfield, S., 1997, \araa, 35, 137
\bibitem[Allard 1998]{Allard98}
Allard, F., 1998, Astronomical Society of the Pacific Conference Series, 134, 370, ed. Rebolo, R., Mart\'\i n, E.L., Zapatero Osorio, M.R.  
\bibitem[Baraffe et al. 1998]{Baraffe98}
Baraffe, I., Chabrier, G., Allard, F., Hauschildt, P.H., 1998, \aap, 337, 403
\bibitem[Basri \& Marcy 1995]{Basri95}
Basri, G., Marcy, G.W., 1995, \aj, 109, 762
\bibitem[Basri et al. 1996]{Basri96}
Basri, G., Marcy, G.W., Oppenheimer, B., Kulkarni, S.R., Nakajima, T., 1996, in Cool Stars, Stellar Systems \& the Sun, 9th Cambridge Workshop, ASP Conference Series, 109, ed. Pallavicini, R. \& Dupree, A.K.
\bibitem[Becklin \& Zuckerman 1988]{Becklin88}
Becklin, E.E. \& Zuckerman, B., 1988, Nature, 336, 656
\bibitem[Burrows et al. 1997]{Burrows97}
Burrows, A., Marley, M., Hubbard, W.B., Lunine, J.I., Guillot, T., Saumon, D., Freedman, R., Sudarsky, D., Sharp, C., 1997, \apj, 491, 856
\bibitem[Burrows et al. 1999]{Burrows99}
Burrows, A., Marley, M., Sharp, C.M., 1999, \apj, submitted
\bibitem[Delfosse 1997]{Delfosse97}
Delfosse, X., 1997, PhD thesis
\bibitem[Delfosse et al. 1998]{Delfosse98}
Delfosse, X., Forveille, T., Perrier, C., Mayor, M., 1998, \aap, 331, 581 
\bibitem[Delfosse et al. 1999]{Delfosse99}
Delfosse, X., Tinney, C.G., Forveille, T., Epchtein, N., Borsenberger, J., Foque, P., Kimeswenger, S., Tiphene, D., 1999, \aaps, 135, 41
\bibitem[deJager \& Nieuwenhuijzen 1987]{deJager87}
DeJager, C. \& Nieuwenhuijzen, H., 1987, \aap, 177, 217
\bibitem[Golimowski et al. 1998]{Golimowski98}
Golimowski, D.A., Burrows, C.J., Kulkarni, S.R., Oppenheimer, B.R., Brukardt, R.A., 1998, \aj, 115, 2579
\bibitem[Goldman et al. 1999]{Goldman99}
Goldman, B., Delfosse, X., Forveille, T., Afonso, C. et al., 1999, \aap preprint 
\bibitem[Gray 1992]{Gray}
Gray, D.F. 1992, {\it The Observation and Analysis of Stellar Photospheres, 2nd Ed.}, New York:Cambridge Univ. Press
\bibitem[Griffith et al. 1998]{Griffith98}
Griffith, C.A., Yelle, R.V., Marley, M.S., 1998, Science, 282, 2063
\bibitem[Jones \& Tsuji 1997]{Jones97}
Jones, H.R.A., Tsuji, T., 1997, \apj, 408, 39
\bibitem[Jones et al. 1994]{Jones94}
Jones, H.R.A., Longmore, A.J., Jameson, R.F., Mountain, C.M., 1994, \mnras, 267, 413
\bibitem[Kirkpatrick et al. 1997]{Kirkpatrick97}
Kirkpatrick, J.D., Henry, T.J., Irwin, M.J., 1997, \aj, 113, 1421
\bibitem[Kirkpatrick et al. 1999a]{Kirkpatrick99a}
Kirkpatrick, J.D. et al., 1999a, \apj, 519, 802

\bibitem[Kirkpatrick et al. 1999b]{Kirkpatrick99b}
Kirkpatrick, J.D. et al., 1999b, \apj, 519, 834
 
\bibitem[Leggett et al. 1998]{Leggett98}
Leggett, S.K., Allard, F., Hauschildt, P.H., 1998, \apj, 509, 836
\bibitem[Mart\'\i n et al. 1997]{Martin97}
Mart\'\i n, E.L., Basri, G., Delfosse, X., Forveille, T., 1997, \aap, 327, L29
\bibitem[Mart\'\i n et al. 1999a]{Martin99a}
Mart\'\i n, E.L., Brandner, W., Basri, G., 1999a, Science, 238, 1718
\bibitem[Mart\'\i n et al. 1999b]{Martin99b}
Mart\'\i n, E.L., Delfosse, X., Basri, G., Goldman, B., Forveille, T., Zapatero Osorio, M.R. 1999b, \aj, 118, 2466

\bibitem[Oppenheimer et al. 1998]{Oppenheimer98}
Oppenheimer, B.R., Kulkarni, S.R., Matthews, K., Van Kerwijk, M.H., 1998, \apj, 502, 932
\bibitem[Rebolo et al. 1998]{Rebolo98}
Rebolo, R., Zapatero Osorio, M.R., Madruga, S., Bejar, V.J.S., Arribas, S., Licandro, J., 1998, Science, 282, 1309
\bibitem[Ruiz et al. 1997]{Ruiz97}
Ruiz, M.T., Leggett, S.K., Allard, F., 1997, \apjl, 491, L110 
\bibitem[Schweitzer et al. 1996]{Schweitzer96}
Schweitzer, A., Hauschildt, P.H., Allard, F., Basri, G., 1996, \mnras, 283, 821
\bibitem[Sharp \& Huebner 1990]{Sharp90}
Sharp, C.M., Huebner, W.F., 1990, \apjs, 72, 417
\bibitem[Tinney \& Reid 1998]{TinneyReid98}
Tinney, C.G., Reid, I.N., 1998, \mnras, 301, 1031
\bibitem[Tinney 1998]{Tinney98a}
Tinney, C.G., 1998, \mnras, 296, L42
\bibitem[Tinney et al. 1998]{Tinney98}
Tinney, C.G., Delfosse, X., Forveille, T., Allard, F., 1998, \aap, 338, 1066
\bibitem[Tokunaga \& Kobayashi 1999]{Tokunaga99}
Tokunaga, A.T., Kobayashi, N., 1999, \aj, 117, 1010
\bibitem[Tsuji et al. 1996a]{Tsuji96a}
Tsuji, T., Ohnaka, K., Aoki, W., Nakajima, T., 1996a, \aap, 308, L29
\bibitem[Tsuji et al. 1996b]{Tsuji96b}
Tsuji, T., Ohnaka, K., Aoki, W., 1996b, \aap, 305, L1 
\bibitem[Tsuji et al. 1999]{Tsuji99}
Tsuji, T., Ohnaka, K., Aoki, W., 1999, \apjl, 520, L119 
\bibitem[Vogt et al. 1994]{Vogt94}
Vogt, S. et al., 1994, in Instrumentation in Astronomy III, Proc. SPIE, 2198, 362, ed. Crawford, D.L., Craine, E.R.

\end{thebibliography}
\end{document}